\documentclass[aps,prd,nofootinbib,onecolumn,a4paper]{revtex4}
\usepackage[colorlinks,linkcolor=blue,anchorcolor=violet,citecolor=red]{hyperref}
\usepackage{amsmath,amssymb}
\usepackage[dvipdfmx]{graphicx}
\usepackage{float}
\usepackage{comment}
\usepackage{xcolor}
\usepackage{bm,latexsym,amsmath,amssymb,amsfonts,mathrsfs}
\usepackage{bbold}
\usepackage{ascmac}
\usepackage{physics}
\usepackage{color,float}
\usepackage{enumerate}

\begin{document}
\title{Quantum uncertainty of gravitational field and entanglement 

in superposed massive particles}%

\author{Yuuki Sugiyama}
 \email{sugiyama.yuki@phys.kyushu-u.ac.jp}
\affiliation{Department of Physics, Kyushu University, 744 Motooka, Nishi-Ku, Fukuoka 819-0395, Japan}

\author{Akira Matsumura}
 \email{matsumura.akira@phys.kyushu-u.ac.jp}
\affiliation{Department of Physics, Kyushu University, 744 Motooka, Nishi-Ku, Fukuoka 819-0395, Japan}
 
\author{Kazuhiro Yamamoto}
 \email{yamamoto@phys.kyushu-u.ac.jp}
\affiliation{Department of Physics, Kyushu University, 744 Motooka, Nishi-Ku, Fukuoka 819-0395, Japan}
\affiliation{
Research Center for Advanced Particle Physics, Kyushu University, 744 Motooka, Nishi-ku, Fukuoka 819-0395, Japan}
\affiliation{
International Center for Quantum-Field Measurement Systems for Studies of the Universe and Particles (QUP), KEK, Oho 1-1, Tsukuba, Ibaraki 305-0801, Japan}

\begin{abstract}
Investigating the quantum nature of gravity is an important issue in modern physics. 
Recently, studies pertaining to the quantum superposition of gravitational potential have garnered significant interest.  
Inspired by Mari \textit{et al.} [Sci. Rep. {\bf 6} 22777 (2016)] and Baym and Ozawa [Proc. Natl. Acad. Sci. U.S.A. {\bf 106}, 3035 (2009)], Belenchia \textit{et al.} [Phys. Rev. D {\bf 98}, 126009 (2018)] considered a gedanken experiment involving such a quantum superposition and mentioned that the superposition renders causality and complementarity inconsistent.
They resolved this inconsistency by considering the quantized dynamical degrees of freedom of gravity.
This suggests a strong relationship between the quantum superposition of the gravitational potential and the quantization of the gravitational field.
In our previous study [Phys. Rev. D {\bf 106}, 125002 (2022)], we have shown that the quantum uncertainty of a field guarantees the consistency between causality and complementarity.
In this study, we focus on the entanglement between two particles' states due to the electromagnetic/gravitational potential and investigate its relationship with quantum uncertainty, causality, and complementarity.
Our numerical analyses show that the quantum uncertainty of the electromagnetic/gravitational field results in vacuum fluctuations and prohibits the entanglement between two particles' states when causality is satisfied.
We further demonstrate that complementarity holds when the particles do not get entangled.
The uncertainty relation does not cause the entanglement between two particles' states, which guarantees complementarity.
\end{abstract}

\maketitle

\section{Introduction}

The unification of gravity and quantum mechanics is essential for understanding extreme phenomena, such as the early universe and the singularity of a black hole, but it is still unclear how the two theories should be unified in a consistent way.
One of the reasons for the difficulty is that any quantum aspects of gravity have not yet been experimentally tested at all.
Therefore, testing the quantum nature of gravity is increasingly important in the development of the quantum theory of gravity.
Authors in Refs.~\cite{Bose2017,Marlleto2017} predicted that two particles, each in a superposition of two spatially localized states, become entangled due to the gravitational potential mediating between the particles.
They suggested that the entanglement originates from the quantum superposition of the gravitational potential, which might exhibit the quantum superposition of the spacetime curvature in the context of general relativity.
The experimental observation of the gravitationally induced entanglement between two particles is expected to be of great help in clarifying the quantum aspects of gravity.
Recently, various studies pertaining to the quantum nature of gravitational potentials have been conducted~(e.g., \cite{Christodoulou,Gnuyen,Anastopoulos,Miki,Aspelmeyer,Schmole,Lopez,Blaushi,Miao,Matsumura,Krisnanda,Datta,Miki2,Eduardo,Kaku1,Miki3,Sugiyama3,Shichijo,Kaku2}), which were inspired by the studies~\cite{Bose2017,Marlleto2017}.
However, the relationship between the quantum nature of the gravitational potential and the quantization of the gravitational field in quantum gravity is debatable.
A gedanken experiment involving the quantum superposition of a massive object, as discussed in Refs.~\cite{Mari,Belenchia2018,Belenchia2019,Danielson2021,Baym,Pesci}, has garnered considerable attention.  
In the gedanken experiment, the quantum superposition of the gravitational potential induced by the object results in inconsistency between causality and complementarity.
This inconsistency is resolved by considering the quantized dynamical degrees of freedom of gravity \cite{Belenchia2018,Belenchia2019,Danielson2021,Pesci}.  
A deep understanding of the gedanken experiment may allow one to clarify the manner by which the quantum nature of the gravitational potential correlates with the quantization of the gravitational field. 

The goal of the present study is to analyze the gedanken experiment similar to that presented in Refs.~\cite{Mari,Belenchia2018,Belenchia2019,Danielson2021,Baym,Pesci} based on quantum electrodynamics (QED)/quantum theory of linearized gravity. 
In this study, we extensively apply the results of our previous studies~\cite{Sugiyama1, Sugiyama2} on electromagnetic fields to a gravitational field within the framework of the linear perturbation theory.
Researchers~\cite{Sugiyama2,Iso1,Iso2} have shown that causality is satisfied by the property of the retarded Green's function. 
Furthermore, the uncertainty relation satisfied for the quantum field has been shown to guarantee the consistency between causality and complementarity~\cite{Sugiyama2}.
In the present study, we focus on the entanglement between two particles due to the electromagnetic/gravitational potential and discuss the role of entanglement in maintaining the consistency.
Our numerical findings show that the uncertainty relation in electromagnetic/gravitational field prohibits the entanglement between particles provided that causality holds. 
In addition, it is shown that the nonentangling feature results in inequality, which implies complementarity.
These clearly demonstrate the importance of quantized electromagnetic/gravitational fields in guaranteeing the consistency between causality and complementarity.

This paper is organized as follows.
In Sec. II, we briefly review the gedanken experiment discussed in Refs.~\cite{Belenchia2018,Belenchia2019,Danielson2021}.
In Sec. III, we present the setup for the electromagnetic/gravitational version of the gedaneken experiment in the present paper.
In Sec. IV, we discuss the relationship between the uncertainty relation of the electromagnetic/gravitational field, the entanglement of two massive particles, and the inequality of complementarity.
Section V presents the summary and conclusions.
Appendix A briefly reviews the QED formulation presented in~\cite{Sugiyama1,Sugiyama2}.
In Appendix B, we demonstrate that the condition \eqref{result} holds.
Throughout the present paper, we adopt the natural units $c=\hbar=1$.
%%%%%%%%%

\section{Brief Review of the gedanken experiment}
\begin{figure}[t]
  \centering
  \includegraphics[width=0.7\linewidth]{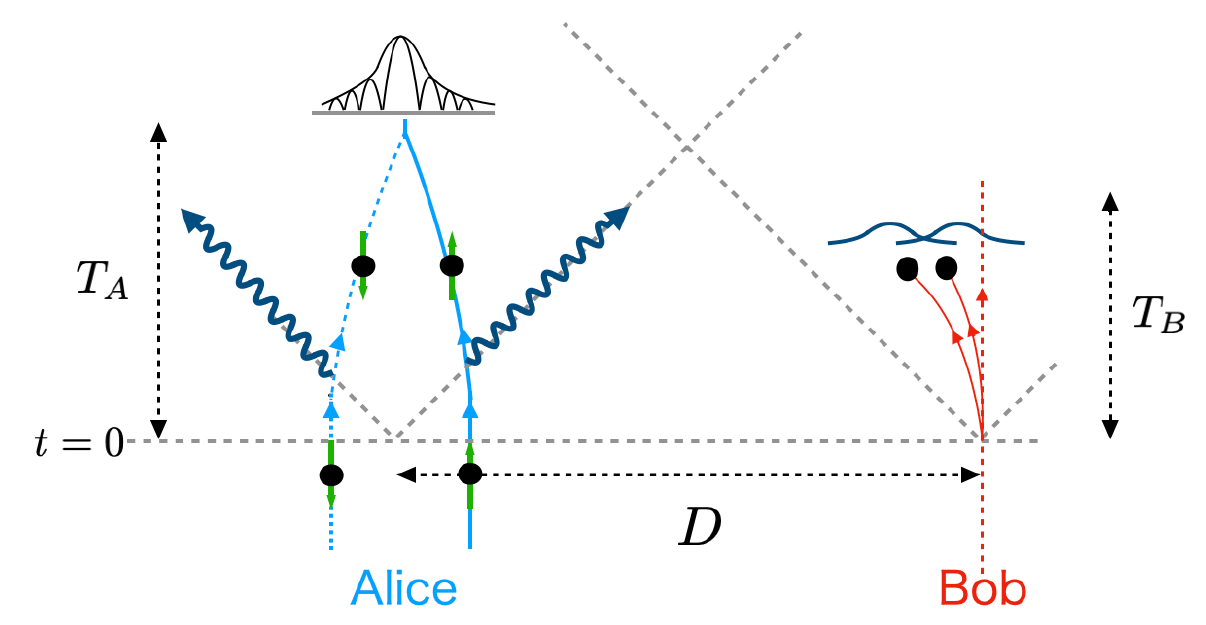}
  \caption{
  Setup for the gedanken experiment.
  $D$ represents the distance between Alice's system and Bob's system. 
  $T_{\text{A}}$ is a time scale for recombining particle A, and $T_{\text{B}}$ is a time scale that particle B in Bob's system will be superposed when he released it.
  Here, we assume $D > T_{\text{A}}$ and $D > T_{\text{B}}$, in which Alice and Bob perform their actions in spacelike separated regions.
  }
  \label{fig:setup}
\end{figure}

We consider two quantum systems (Alice's particle and Bob's particle) separated by a distance $D$ interacting through the electromagnetic/gravitational potential (Fig.~\ref{fig:setup}).
In Alice's system, her particle is prepared in a quantum superposition of two locations and starts to recombine during time $T_{\text{A}}$.
At $t=T_{\text{A}}$, Alice performs an interference experiment and assesses whether it will be successful (whether the interference pattern of her particle will be observed).
If the superposition state of Alice's particle is preserved, then the interference experiment will be successful; however, if the superposition state is not preserved, then the experiment will not be  successful.
In Bob's system, Bob chooses whether he releases his particle or not at $t=0$.
When he releases his particle, it is affected by the electromagnetic/gravitational potential due to Alice's particle and is thus displaced. 
Because Alice's particle is in the superposition of the two paths, the magnitude of the potential perceived by Bob's particle changes depending on the path traversed by her particle. 
Thus, Bob can use his particle to measure which path Alice’s particle took.

Let us assume that Alice's interference experiment during the time
$T_\text{A}$ 
and Bob's choice and measurement during the time
$T_\text{B}$ 
are performed in a spacelike separated region satisfying 
$D>T_\text{A}$ 
and 
$D>T_\text{B}$ 
(Fig.~\ref{fig:setup}).
If Bob releases his particle and can measure the position of Alice's particle, then, by complementarity, the superposition state of Alice's particle collapses and the particle decoheres. 
Thus, the interference experiment is not successful.
By contrast, when Bob decides not to release his particle and does not measure the path undertaken by Alice’s particle, then her particle will preserve the superposition state and her interference experiment will be successful.
This indicates that causality is violated because Bob's choice is known by Alice when her particle is in a region where his actions have no influence causally.
However, if the causality holds, then Alice's interference experiment is successful (she observes the interference pattern of her particle). 
In this case, without decohering Alice's particle, Bob can use his released particle to obtain the which-path information of her particle.
This results in a violation in complementarity.
%%%%
%%%%
The inconsistency between causality and complementarity can be resolved by considering the vacuum fluctuations of the electromagnetic/gravitational field and the emissions of photons/gravitons, which was demonstrated as an order estimation in Refs.~\cite{Belenchia2018,Belenchia2019,Pesci}.
Bob’s measurement to acquire the which-path information of Alice's particle is limited by the vacuum fluctuations of a quantized electromagnetic/gravitational field, and Alice’s interference experiment fails because of the decoherence induced by the entangling radiation of photons/gravitons.
Here, the entangling radiation refers to the radiation emitted from and entangled with Alice's particle.
This suggests that a quantized electromagnetic/gravitational field is sufficient to avoid the inconsistency between causality and complementarity.

In our previous study~\cite{Sugiyama2}, we discovered that the uncertainty relation represented by Robertson's inequality for a quantized field guarantees the consistency between causality and complementarity.  
In the following sections, based on the QED/quantum theory of linearized gravity, we provide a detailed discussion regarding consistency by focusing on the entanglement between two charged/massive particles.

\section{Analysis of the QED/quantum theory of linearized gravity version}
In this section, we consider the electromagnetic and gravitational versions of a similar gedanken experiment based on QED and the quantum theory of linearized gravity, respectively.
In our analysis, we consider two charged/massive particles, A and B, which are nonrelativistic and obey the framework of quantum mechanics; the electromagnetic/gravitational field coupled to the particles is assumed to be a quantum field. 
\begin{figure}[htbp]
  \includegraphics[width=0.6\linewidth]{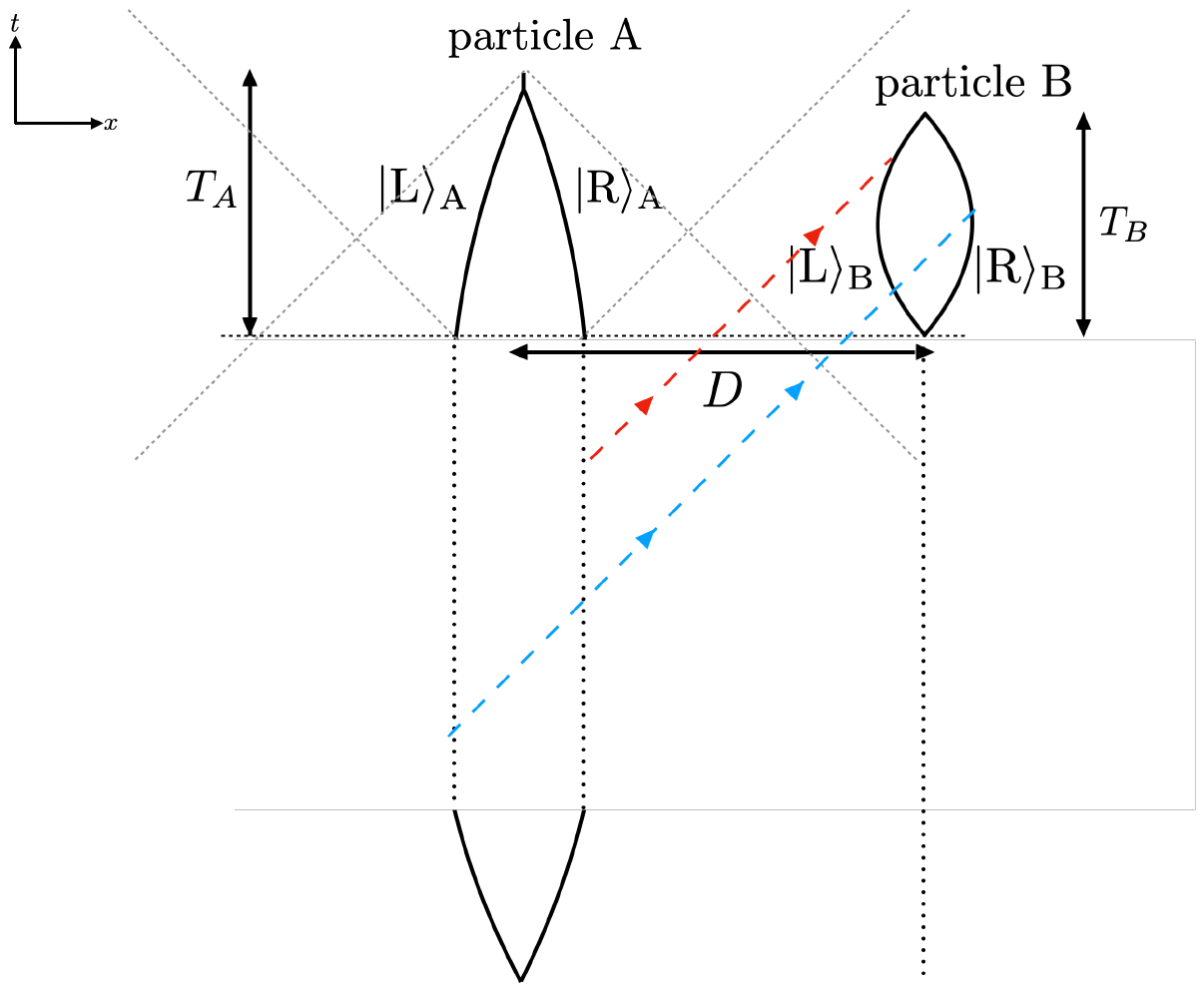}
\caption
{
 Configuration of our model.
 We specify regimes $D  > T_\text{A}$ and $D>T_\text{B}$, in which the retarded Green's function propagating from particle B to A vanishes.
 Particle A traverse via the right or left path $|\text{R}\rangle_{\text{A}}$($|\text{L}\rangle_{\text{A}}$) and induces an electromagnetic/gravitational field along each path (as shown by the dashed red or blue line). 
 The retarded field caused by particle A affects particle B traversing via the left ($|\text{L}\rangle_{\text{B}}$) or right ($|\text{R}\rangle_{\text{B}}$) path.
  \label{fig:configuration}
}
\end{figure}
%%%%
We focus on the similarity between the electromagnetic field and gravitational fields, and extend the results obtained from the analysis of an electromagnetic field in a previous study~\cite{Sugiyama1, Sugiyama2} to a gravitational field.
In this extension, we introduce several important assumptions.
We consider a linearized regime of gravity by expanding the metric of spacetime around the Minkowski spacetime metric $\eta_{\mu\nu}$.
The full spacetime metric is given by $g_{\mu\nu}=\eta_{\mu\nu}+h_{\mu\nu}$, where $h_{\mu\nu}$ is the metric perturbation satisfying $|h_{\mu\nu}|\ll1$.
\footnote{
Let us now consider the gravitational interaction of two particles with the same masses $m$.
Based on the linearized gravity theory~\cite{Wheeler}, the energy-momentum tensor of a particle $T_{\mu\nu}$ induces the fluctuation component of the metric 
$h_{\mu\nu}
\sim
G\int d^3y {T_{\mu\nu}(t_{\text{r}}, \bm{y})}/{|\bm{x}-\bm{y}|}$
with the gravitational constant $G$.
Here, $t_{\text{r}}=t-|\bm{x}-\bm{y}|$ is the retarded time, which represents the delay with respect to the propagation from the source point $\bm{y}$ to a spacetime point $\bm{x}$.
The components of $h_{\mu\nu}$ are evaluated as
$h_{00} \sim G{m}/{R},
\quad
h_{0i} \sim h_{00} ({L}/{T})\bm{e}_{i}$
, and
$h_{ij} \sim h_{00} ({L}/{T})^2\bm{e}_{i}\bm{e}_{j}$,
where $T$ is the total time scale when particle A is spatially superposed, and $L$ is the separation of spatial superposition of particles A and B.
$R$ characterizes the typical size of each particle satisfying $R \lesssim L$.
$(L/T)\bm{e}_{i}$ denotes the characteristic velocity of the system in the $i$ direction of the unit vector $\bm{e}_{i}$.
Considering the nonrelativistic condition $L/T \ll 1$, the condition $|h_{\mu\nu}|\ll1$ is valid when $G{m}/{L}
\ll1$ is satisfied.
}
The initial state of the particles is assumed to be each in spatially localized superposition (Fig.~\ref{fig:configuration}), which might be realized via the Stern-Gerlach effect, as explained in Appendix A.
Additionally, we assume that no initial entanglement occurs between the particles and the electromagnetic/gravitational field.
\footnote{
In the case of gravity, Alice’s particle may be entangled with her apparatus because of the conservation of energy-momentum.
For a rigorous description, we must consider the effects induced in a laboratory, as discussed in~\cite{Suzuki}.
However, the authors of~\cite{Belenchia2018} argued that laboratory effects need not be considered because the state of the laboratory does not produce a significant decoherence.
The authors of~\cite{Howl,Howl2} discussed the effect of the gravitational potential arising from the apparatus in a laboratory on the relative phases that cause entanglement between two particles.
If the laboratory apparatus is sufficiently heavy, then it does not shift significantly.
Thus the apparatus will not be superposed state and not contribute to the relative phases.
Moreover, if the gravitational field created by the apparatus is homogeneous, then the phase shift due to its field will also not affect the relative phase.
} 
Furthermore, the wave packets of these particles are sufficiently far apart to form local paths within each device.

Then, we can define the current of particle A (B) as ${J}^{\mu}_{\text{AP}}(x)$ (${J}^{\mu}_{\text{BQ}}(x)$) localized around their paths of P (Q)$(=R, L)$.
Under the assumptions above, as shown in Appendix A, the decoherence and the entanglement between the two particles for
the electromagnetic version can be described by the following quantities ~\eqref{gammaiEM}$-$\eqref{PhiEM}
\begin{align}
\Gamma^{\text{EM}}_{i}
&=
\frac{1}{4}\int d^4x\int d^4y 
\Delta J^{\mu}_{i}(x)\Delta J^{\nu}_{i}(y)
\langle 
\{\hat{A}^\text{I}_{\mu}(x), \hat{A}^\text{I}_{\nu}(y)\}
\rangle
=
\frac{1}{4}\langle\{\hat{\phi}^{\text{EM}}_{i},\hat{\phi}^{\text{EM}}_{i}\}\rangle
=
\frac{1}{2}\langle0|(\hat{\phi}^{\text{EM}}_{i})^2|0\rangle,
\label{gammaiEM}
\\
\quad
\Gamma^{\text{EM}}_\text{c}
&=
\frac{1}{2}\int d^4x\int d^4y
\Delta J^{\mu}_{\text{A}}(x)\Delta J^{\nu}_{\text{B}}(y)
\langle 
\{\hat{A}^\text{I}_{\mu}(x), \hat{A}^\text{I}_{\nu}(y)\}
\rangle
=
\frac{1}{2}\langle\{\hat{\phi}^{\text{EM}}_{\text{A}},\hat{\phi}^{\text{EM}}_{\text{B}}\}\rangle,
\label{gammacEM}
\\
\quad
\Phi^{\text{EM}}
&=
\frac{1}{2}\int d^4x d^4y
\Big\{
\Delta J^{\mu}_{\text{A}}(x)\Delta J^{\nu}_{\text{B}}(y)
+
\Delta J^{\mu}_{\text{B}}(x)\Delta J^{\nu}_{\text{A}}(y)
\Big\}
G^{\text{r}}_{{\mu\nu}}(x,y)
=
\frac{1}{2}(\Phi^{\text{EM}}_{\text{AB}}+\Phi^{\text{EM}}_{\text{BA}}),
\label{PhiEM}
\end{align}
where we defined $\Delta J^{\mu}_{i}=J^{\mu}_{i\text{R}}-J^{\mu}_{i\text{L}}$  with $i=\text{A}, \text{B}$.
Similarly, we can define the energy-momentum tensors of particle A (B) as ${T}^{\mu \nu}_{\text{AP}}(x)$ (${T}^{\mu \nu}_{\text{BQ}}(x)$) localized around their paths of P (Q)$(=R, L)$, and the gravitational field version is presented as follows:
\begin{align}
\Gamma^{\text{GR}}_{i}
&=
\frac{1}{4}\int d^4x\int d^4y 
\Delta T^{\mu\nu}_{i}(x)\Delta T^{\rho\sigma}_{i}(y)
\langle 
\{\hat{h}^\text{I}_{\mu\nu}(x), \hat{h}^\text{I}_{\rho\sigma}(y)\}
\rangle
=
\frac{1}{4}\langle\{\hat{\phi}^{\text{GR}}_{i},\hat{\phi}^{\text{GR}}_{i}\}\rangle
=
\frac{1}{2}\langle0|(\hat{\phi}^{\text{GR}}_{i})^2|0\rangle,
\label{gammaig}
\\
\quad
\Gamma^{\text{GR}}_\text{c}
&=
\frac{1}{2}\int d^4x\int d^4y
\Delta T^{\mu\nu}_{\text{A}}(x)\Delta T^{\rho\sigma}_{\text{B}}(y)
\langle 
\{\hat{h}^\text{I}_{\mu\nu}(x), \hat{h}^\text{I}_{\rho\sigma}(y)\}
\rangle
=
\frac{1}{2}\langle\{\hat{\phi}^{\text{GR}}_{\text{A}},\hat{\phi}^{\text{GR}}_{\text{B}}\}\rangle,
\label{gammacg}
\\
\quad
\Phi^{\text{GR}}
&=
\frac{1}{2}\int d^4x d^4y
\Big\{
\Delta T^{\mu\nu}_{\text{A}}(x)\Delta T^{\rho\sigma}_{\text{B}}(y)
+
\Delta T^{\mu\nu}_{\text{B}}(x)\Delta T^{\rho\sigma}_{\text{A}}(y)
\Big\}
G^{\text{r}}_{{\mu\nu\rho\sigma}}(x,y)
=
\frac{1}{2}(\Phi^{\text{GR}}_{\text{AB}}+\Phi^{\text{GR}}_{\text{BA}}),
\label{phig}
\end{align}
where we defined $\Delta T^{\mu\nu}_{i}=T^{\mu\nu}_{i\text{R}}-T^{\mu\nu}_{i\text{L}}$ with $i=\text{A}, \text{B}$.
The operators $\hat{\phi}^{\text{EM}}_{i}$ and $\hat{\phi}^{\text{GR}}_{i}$ describe the phase shifts due to the quantum fluctuations of the electromagnetic field and gravitational field, respectively, which are expressed as
\begin{align}
\hat{\phi}^{\text{EM}}_{i}
=
\int d^4x 
\Delta J^{\mu}_{i}(x)\hat{A}^{\text{I}}_{\mu}(x),
\quad
\hat{\phi}^{\text{GR}}_{i}
=
\int d^4x 
\Delta T^{\mu\nu}_{i}(x)\hat{h}^{\text{I}}_{\mu\nu}(x).
\label{phiemg}
\end{align}
Here, $\langle \bigl\{\hat{A}^\text{I}_{\mu}(x), \hat{A}^\text{I}_{\mu}(y)\bigr\}\rangle$ 
and 
$G^{\text{r}}_{\mu\nu}(x,y)$ are the two-point function of the vacuum state $|0\rangle$ and the retarded Green's function with respect to the quantized electromagnetic field  in the interaction picture, while $\langle \bigl\{\hat{h}^\text{I}_{\mu\nu}(x), \hat{h}^\text{I}_{\mu\nu}(y)\bigr\}\rangle$
and $G^{\text{r}}_{\mu\nu\rho\sigma}(x,y)$~ are the version for 
the gravitational field \cite{Donoghue}.
The quantities $\Phi^{\text{EM}}_{\text{AB}}$ and $\Phi^{\text{EM}}_{\text{BA}}$ are defined as
\begin{align}
\Phi^{\text{EM}}_{\text{AB}}
&=
\int d^4x d^4y
\Delta J^{\mu}_{\text{A}}(x)\Delta J^{\nu}_{\text{B}}(y)G^{\text{r}}_{\mu\nu}(x,y),
\quad
\Phi^{\text{EM}}_{\text{BA}}
=
\int d^4x d^4y
\Delta J^{\mu}_{\text{B}}(x)\Delta J^{\nu}_{\text{A}}(y)G^{\text{r}}_{\mu\nu}(x,y).
\end{align}
The gravitational version of the quantities $\Phi^{\text{GR}}_{\text{AB}}$ and $\Phi^{\text{GR}}_{\text{BA}}$ can be similarly expressed as
\begin{align}
\Phi^{\text{GR}}_{\text{AB}}
&=
\int d^4x d^4y
\Delta T^{\mu\nu}_{\text{A}}(x)\Delta T^{\rho\sigma}_{\text{B}}(y)G^{\text{r}}_{\mu\nu\rho\sigma}(x,y),
\quad
\Phi^{\text{GR}}_{\text{BA}}
=
\int d^4x d^4y
\Delta T^{\mu\nu}_{\text{B}}(x)\Delta T^{\rho\sigma}_{\text{A}}(y)G^{\text{r}}_{\mu\nu\rho\sigma}(x,y).
\end{align}

In the following, we present the inequality representing complementarity, the uncertainty relation, and one of the entanglement measure: negativity.
The inequality, the uncertainty relation, and the negativity for the electromagnetic case are evaluated from the quantum state of the charged particles determined using $\Gamma^\text{EM}_{i} \, (i=\text{A},\text{B})$, 
$\Gamma^\text{EM}_{\text{c}}$, 
$\Phi^\text{EM}_{\text{AB}}$ and 
$\Phi^\text{EM}_{\text{BA}}$ (See Appendix A). 
By replacing these quantities with 
$\Gamma^\text{GR}_{i} \, (i=\text{A},\text{B})$, 
$\Gamma^\text{GR}_{\text{c}}$, 
$\Phi^\text{GR}_{\text{AB}}$ and $\Phi^\text{GR}_{\text{BA}}$, we obtain the formulas for gravitational case.
Subsequently, we adopt simple notations
$\Gamma_i\, (i=\text{A},\text{B})$, 
$\Gamma_\text{c}$, 
$\Phi_\text{AB}$ 
and 
$\Phi_\text{BA}$ to describe the quantities above for the electromagnetic and gravitational cases in a unified manner.

We first introduce the visibility $\mathcal{V}_{\text{A}}$ of particle A and distinguishability $\mathcal{D}_{\text{B}}$ (the which-path information of particle A acquired from particle B). 
These two quantities are useful for expressing complementarity. 
Based on previous studies~\cite{Jaeger, Englert}, complementarity holds if the visibility $\mathcal{V}_{\text{A}}$ and distinguishability $\mathcal{D}_{\text{B}}$ satisfy the following inequality: 
\begin{align}
\mathcal{V}^2_{\text{A}}+\mathcal{D}^2_{\text{B}} \leq 1.
\end{align}
Using the results of~\eqref{inequalityEM} and~\eqref{inequalityg}, the complementarity is formulated as follows:
\begin{align}
\mathcal{V}^2_{\text{A}}+\mathcal{D}^2_{\text{B}}
=
e^{-2\Gamma_{\text{A}}}\cos^2\left(\frac{\Phi_{\text{AB}}}{2}\right)
+
e^{-2\Gamma_{\text{B}}}\sin^2\left(\frac{\Phi_{\text{BA}}}{2}\right)
\leq 1. 
\end{align}

Next, we consider the Schr\"{o}dinger-Robertson uncertainty relation as follows:
\begin{align}
(\Delta {\phi}_{\text{A}})^2(\Delta {\phi}_{\text{B}})^2
\geq
\frac{1}{4}\Big(\langle \{\hat{\phi_{\text{A}}}, \hat{\phi}_{\text{B}}\}\rangle\Big)^2
+
\frac{1}{4}\Big|\langle[\hat{\phi}_{\text{A}}, \hat{\phi}_{\text{B}}]\rangle\Big|^2,
\end{align}
where 
$(\Delta \phi_{i})^2=\langle0|\hat{\phi}^2_{i}|0\rangle-(\langle0|\hat{\phi}_{i}|0\rangle)^2$
($i=\text{A}, \text{B}$) 
are the variances of the operators 
$\hat{\phi}_{i}=\hat{\phi}^\text{EM}_i, \hat{\phi}^\text{GR}_i$ 
defined in Eq.~\eqref{phiemg}.
Subsequently, the following Schr\"{o}dinger-Robertson uncertainty relation can be obtained:
\begin{align}
\Gamma_{\text{A}}\Gamma_{\text{B}}
&\geq
\frac{\Gamma^2_{\text{c}}}{4}
+
\frac{1}{16}(\Phi_{\text{AB}}-\Phi_{\text{BA}})^2,
\end{align}
where we used 
$(\Delta \phi_{i})^2=\langle0|\hat{\phi}^2_{i}|0\rangle-(\langle0|\hat{\phi}_{i}|0\rangle)^2=2\Gamma_{i}$
and 
$\langle \{\hat{\phi_{\text{A}}}, \hat{\phi}_{\text{B}}\}\rangle=2\Gamma_{\text{c}}$, 
followed by Eqs.~\eqref{gammaiEM}, \eqref{gammaig} and \eqref{gammacEM}, \eqref{gammacg}.
The expectation value of the commutator, 
$\langle[\hat{\phi}_{\text{A}}, 
\hat{\phi}_{\text{B}}]\rangle=i(\Phi_{\text{AB}}-\Phi_{\text{BA}})$, 
is obtained from the same calculation as for Eq.~\eqref{commutation}.
The inequality above shows that the product of 
$\Gamma_{\text{A}}$ and 
$\Gamma_{\text{B}}$ has a lower bound expressed by 
$\Gamma_\text{c}$, $\Phi_\text{AB}$, and $\Phi_\text{BA}$.

Finally, we present the formula for negativity $\mathscr{N}$ \cite{Vidal2002}.
The negativity is convenient to determine whether two particles A and B are entangled or not.
Let us consider a density matrix 
$\rho$ of a bipartite system AB. 
The negativity is defined as follows: 
\begin{align}
\mathscr{N}=\sum_{\lambda_i<0} |\lambda_{i}|,
\end{align}
where $\lambda_{i}$ represent the negative eigenvalues of the partial transposition 
$\rho^{\text{T}_{\text{A}}}$ with the elements 
$\langle a| \langle b| \rho^{\text{T}_{\text{A}}}|a' \rangle |b'\rangle=\langle a'| \langle b| \rho |a \rangle |b'\rangle$ in a basis 
$\{|a \rangle |b \rangle \}_{a,b}$ of the system AB. 
Specifically, only one negative eigenvalue $\lambda_{\text{min}}$ of the partial transposed density matrix of a two-qubit system exists~\cite{Sanpera1998}.
In our system, the quantum state of two particles A and B are regarded as a two-qubit state since each of particle is in a superposition of the left path ($|\text{L} \rangle_\text{A(B)}$ ) or right path ($|\text{R} \rangle_\text{A(B)}$), as shown in Fig.\ref{fig:configuration}.
Thus, we can rewrite the negativity as $\mathscr{N}=\max[-\lambda_\text{min},0]$.
If $\mathscr{N}=0$ or $\lambda_{\text{min}} \geq 0$ holds, then the two particles are not entangled. 
The minimum eigenvalue is obtained as follows:
\begin{align}
\lambda_{\text{min}}
&=
\frac{1}{4}
\Big[
1-e^{-\Gamma_{\text{A}}-\Gamma_{\text{B}}} \cosh[\Gamma_\text{c}]- 
\Big\{
\big(e^{-\Gamma_{\text{A}}}-e^{-\Gamma_{\text{B}}}\big)^2+4e^{-\Gamma_{\text{A}}-\Gamma_{\text{B}}}\sin^2\Big[\frac{(\Phi_{\text{AB}}+\Phi_{\text{BA}})}{4}\Big]+e^{-2\Gamma_{\text{A}}-2\Gamma_{\text{B}}} \sinh^2[\Gamma_\text{c}]
\Big
\}^{\frac{1}{2}}
\Big].
\label{negativity}
\end{align}
In the next section, we discuss the relationship among the uncertainty relation of the electromagnetic/gravitational fields, the inequality representing complementarity, and the quantity ${\lambda}_{\text{min}}$.

\section{Role of entanglement on uncertainty relation of field and complementarity}
\begin{figure}[t]
  \centering
  \includegraphics[width=0.6\linewidth]{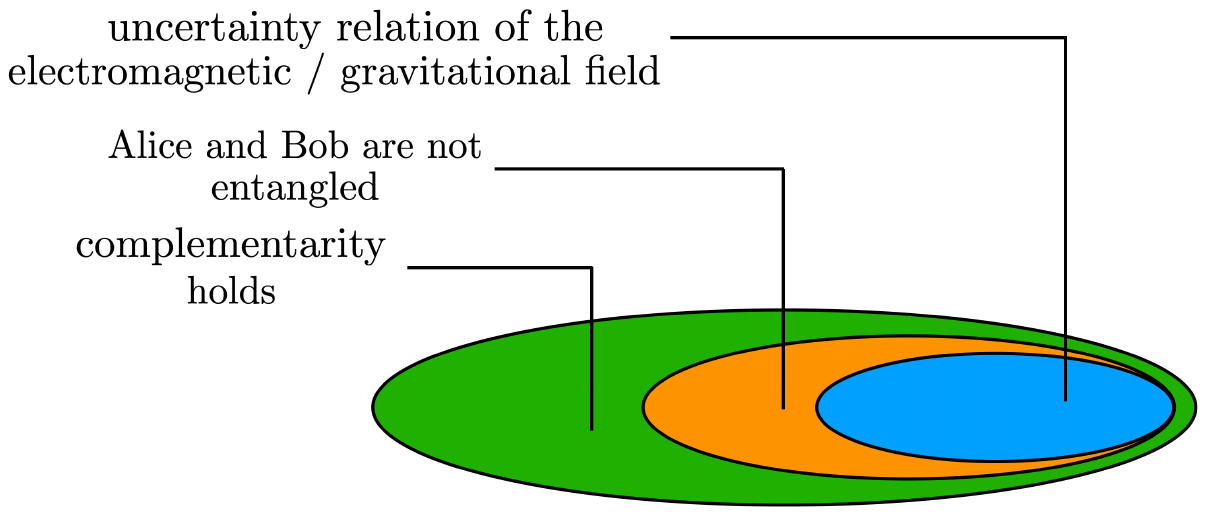}
  \caption{
  Inclusion relationship of the uncertainty relation (blue region), the condition of the nongeneration of entanglement (orange region), and the complementarity inequality (green region) are inclusive.
  }
  \label{fig:inclusive}
\end{figure}

In this section, we reveal how the uncertainty relation relates to complementarity, using the entanglement between two particles A and B.
In our gedanken experiment, we consider the region where Bob's effect does not propagate to Alice's system.
Using the retarded Green's function, which describes the causal influence of a source, this is quantified as $\Phi_{\text{AB}}=0$~\cite{Sugiyama1,Sugiyama2,Iso1,Iso2}.
This result reflects a general property of the retarded Green's function, which holds for both electromagnetic and gravitational fields.
%This result is a general property of the retarded Green's function that holds true for both electromagnetic and gravitational field.
Therefore, the complementarity inequality, Schr\"{o}dinger-Robertson uncertainty relation, and $\lambda_{\text{min}}$ are expressed as follows, respectively:
%Then complementarity inequality, Schr\"{o}dinger-Robertson uncertainty relation, and $\lambda_{\text{min}}$ become
\begin{align}
e^{-2\Gamma_{\text{A}}}
&+
e^{-2\Gamma_{\text{B}}}\sin^2\left(\frac{\Phi_{\text{BA}}}{2}\right)
\leq 1,
\label{inequalityapp}
\\
\quad
\Gamma_{\text{A}}\Gamma_{\text{B}}
&\geq
\frac{\Gamma^2_{\text{c}}}{4}+\frac{\Phi^2_{\text{BA}}}{16},
\label{robertsonapp}
\\
\quad
\lambda_{\text{min}}
=
\frac{1}{4}
\Big[
1-e^{-\Gamma_{\text{A}}-\Gamma_{\text{B}}} \cosh[\Gamma_\text{c}]- 
&
\Big\{
\big(e^{-\Gamma_{\text{A}}}-e^{-\Gamma_{\text{B}}}\big)^2+4e^{-\Gamma_{\text{A}}-\Gamma_{\text{B}}}\sin^2\Big[\frac{\Phi_{\text{BA}}}{4}\Big]+e^{-2\Gamma_{\text{A}}-2\Gamma_{\text{B}}} \sinh^2[\Gamma_\text{c}]
\Big
\}^{\frac{1}{2}}
\Big].
\label{negativityapp}
\end{align}
In our recent study \cite{Sugiyama2}, we discovered that the inequality presented in \eqref{robertsonapp} is the sufficient condition for the complementarity inequality [Eq.~\eqref{inequalityapp}] in the electromagnetic case [strictly, we used the Robertson uncertainty relation 
$\Gamma_{\text{A}}\Gamma_{\text{B}}\geq \Phi^2_{\text{BA}}/16$, which follows by~\eqref{robertsonapp}].
To reveal the relationship between the inequality~\eqref{robertsonapp} 
and complementarity inequality~\eqref{inequalityapp}, 
we consider the role of entanglement.
We first focus on the relationship between the uncertainty relation of the electromagnetic/gravitational field [Eq.~\eqref{robertsonapp}]
and
$\lambda_{\text{min}}$.
Let us consider the limit of small coupling constants for the electromagnetic/gravitational cases. 
The quantities 
$\Gamma_{i}$, 
$\Gamma_{\text{c}}$, 
and $\Phi_{\text{BA}}$ depending on the coupling constants are small, and the approximate form of 
$\lambda_{\text{min}}$ can be expressed as
\begin{align}
\lambda_{\text{min}}
\approx
\frac{1}{4}
\left[
\Gamma_{\text{A}}+\Gamma_{\text{B}}
-
\sqrt{
\big(
\Gamma_{\text{A}}+\Gamma_{\text{B}}
\big)^2
-4
\Big(
\Gamma_{\text{A}}\Gamma_{\text{B}}
-\frac{\Gamma^2_{\text{c}}}{4}
-\frac{\Phi^2_{\text{BA}}}{16}
\Big)}
\ 
\right],
\label{inequalityappb}
\end{align}
which is valid for 
$\Gamma_{i} \ll1$ ($i=\text{A}, \text{B}$), $|\Gamma_{\text{c}}|\leq\Gamma_{\text{A}}+\Gamma_{\text{B}}\ll1$, and $|\Phi_{\text{BA}}|\ll1$.
The inside of the square root is always positive because of $(\Gamma_{\text{A}}+\Gamma_{\text{B}})^2-4(\Gamma_{\text{A}}\Gamma_{\text{B}}-\Gamma^2_{\text{c}}/4-\Phi^2_{\text{BA}}/16)=(\Gamma_{\text{A}}-\Gamma_{\text{B}})^2+\Gamma^2_{\text{c}}+\Phi^2_{\text{BA}}/4\geq0$. 
On the other hand, the sign of $\Gamma_{\text{A}}\Gamma_{\text{B}}-\Gamma^2_{\text{c}}/4-\Phi^2_{\text{BA}}/16$ inside of the square root in Eq.~\eqref{inequalityappb} determines the sign of $\lambda_{\rm min}$, i.e., 
appearance of entanglement between two particles.
From the Schr\"{o}dinger-Robertson uncertainty relation~\eqref{robertsonapp},  
$\Gamma_{\text{A}}\Gamma_{\text{B}}-\Gamma^2_{\text{c}}/4-\Phi^2_{\text{BA}}/16$ must be non-negative. 
Therefore, the Schr\"{o}dinger-Robertson uncertainty relation and the entanglement between the particles A and B appear to be correlated.
This observation, which is obtained on the basis of the approximation, is 
extended to more general relationship among the complementarity inequality, Schr\"odinger-Robertson uncertainty relation, and the nonentanglement property between the two particles. 
Namely, we can demonstrate the following relationship of the sufficient conditions numerically (for a more detailed explanation, see Appendix B):
\begin{align}
\Gamma_{\text{A}}\Gamma_{\text{B}}
\geq
\frac{\Gamma^2_{\text{c}}}{4}
+\frac{\Phi^2_{\text{BA}}}{16}
\quad
\Longrightarrow
\quad
{\lambda}_{\text{min}}\geq0
\quad
\Longrightarrow
\quad
e^{-2\Gamma_{\text{A}}}
+
e^{-2\Gamma_{\text{B}}}\sin^2\left(\frac{\Phi_{\text{BA}}}{2}\right)
\leq 1.
\label{result}
\end{align}
%%%%
The relationship of the above sufficient conditions~\eqref{result} are depicted in Fig.~\ref{fig:inclusive}, which are obtained under the causality
condition that the Bob's action is spacelike separated from Alice, i.e., $\Phi_{\text{AB}}=0$. 
The relationship presented in \eqref{result} 
mean as follows:
The Schr\"odinger-Robertson uncertainty relation implies the existence of the vacuum fluctuations of electromagnetic/gravitational field because 
$\Gamma_{\rm A}$ and $\Gamma_{\rm B}$ must be nonzero since $\Phi_{\rm BA}$ is nonzero.  
The origin of the nonzero values of $\Gamma_{\rm A}$ and $\Gamma_{\rm B}$ is the decoherence of the superposition of each particle, which is supposed to come from  
the entanglement between the particles and the electromagnetic/gravitational field.
This entanglement causes no generation of the entanglement between particles A and B, i.e., $\lambda_{\rm min}\geq0$.
Because particles A and B are not entangled, Bob is not able to sufficiently get the which-path information of Alice's particle.
Therefore, the complementarity inequality holds. 

\section{Conclusion}

The gedanken experiment discussed in Refs.~\cite{Mari,Belenchia2018,Belenchia2019,Danielson2021,Baym} indicated that the quantum superposition of gravitational potential may result in inconsistency between causality and complementarity.
The authors of~\cite{Belenchia2018,Belenchia2019,Danielson2021} argued the inconsistency is resolved by the vacuum fluctuations and entangling radiation 
of electromagnetic/gravitational field.
In the present study, based on the QED/quantum theory of linearized gravity, we analysed the gedanken experiment in connection with the Schr\"odinger-Robertson uncertainty relation, complementarity inequality, focusing on the entanglement between two particles. 
We have discovered that the Schr\"odinger-Robertson uncertainty relation in the electromagnetic/gravitational field prohibits the generation of entanglement between two particles when the causality is fulfilled, which is consistent with the result of our previous study~\cite{Sugiyama1}.
Additionally, we have numerically demonstrated that the condition, under which two particles are not entangled, guarantees complementarity.
The essence of this inconsistency is the assumption that the entanglement between Alice and Bob occurs in the region where Bob's information cannot causally propagate to Alice.
Our results have shown that the two particles can not be entangled because of the quantized electromagnetic/gravitational field, which resolved the paradox and preserved the consistency between causality and complementarity.

Thus, the essence of the resolution for paradox in this gedanken experiment is the existence of the vacuum fluctuation and entangling radiation, which cause the decoherence.
This decoherence is supposed to be induced by the entanglement between the particle and field.
However, determining whether a particle and a field are entangled or not is nontrivial. 
It will be important to discuss the condition that the particle and field are generally entangled.
Furthermore, the structure of the entanglement between particle and field can be further investigated using various quantities of quantum information.
These studies are left for future work.

\acknowledgements
We thank Bei-Lok Hu, Satoshi Iso, and Yasusada Nambu for useful discussions and comments.
Y.S. was supported by the Kyushu University Innovator Fellowship in Quantum Science.
A.M. was supported by JSPS KAKENHI (Grants No.~JP23K13103 and No.~JP23H01175).
K.Y. was supported by JSPS KAKENHI (Grants No.~JP22H05263 and No.~JP23H01175).

\begin{appendix}
\section{Summary of QED formulation}
The discussion presented herein of the gravitational field is analogous to the results obtained in our previous study~\cite{Sugiyama1,Sugiyama2}.
Here, we briefly summarize the formulation of the QED.
We introduce a model of two charged particles, A and B, coupled to an electromagnetic field.
The total Hamiltonian of the proposed system comprises the local Hamiltonians of charged particles 
$\hat{H}_{\text{A}}$ and $\hat{H}_{\text{B}}$, the free Hamiltonian of the quantized electromagnetic field 
$\hat{H}_\text{EM}$, and the interaction term $\hat{V}$ as
%%%%
\begin{eqnarray}
\hat{H}=\hat{H}_{\text{A}}+\hat{H}_{\text{B}}+\hat{H}_\text{EM}+\hat{V}, 
\quad 
\hat{V}=\int d^3x \Big(\hat{J}^\mu_{\text{A}}(\bm x)+\hat{J}^\mu_{\text{B}}(\bm{x})\Big)\hat{A}_{\mu}(\bm{x}),
\end{eqnarray}
%%%%
where 
$\hat{J}^\mu_{\text{A}} $ 
and 
$\hat{J}^\mu_{\text{B}} $ are the current operators of each particle coupled with the gauge field operator $\hat{A}_\mu $.
We consider that each particle is in a superposition of two trajectories
\begin{align}
|\Psi(0)\rangle
&=
\frac{1}{{2}}
(|\text{L}\rangle_{\text{A}}+|\text{R}\rangle_{\text{A}})
(|\text{L}\rangle_{\text{B}}+|\text{R}\rangle_{\text{B}})
|\alpha\rangle_{\text{EM}},
\end{align}
which is described by the localized states 
$|\text{L} \rangle_j$ and 
$|\text{R} \rangle_j$
of particle 
$j=\text{A},\text{B}$. 
The electromagnetic field is in a coherent state $|\alpha\rangle_{\text{EM}}$ with $|\alpha \rangle_\text{EM}=\hat{D}(\alpha)|0\rangle_\text{EM}$.
$|0\rangle_\text{EM}$ is the vacuum state satisfying
$\hat{a}_\mu (\bm{k})|0 \rangle_\text{EM}=0$ for the annihilation operator of the electromagnetic field $\hat{a}_\mu (\bm{k})$; and 
$\hat{D}(\alpha)$ is the unitary operator known as a displacement operator, which is defined as
%%%%
\begin{equation}
\hat{D}(\alpha)=\exp \left[
\int d^3k (\alpha^\mu (\bm{k})\hat{a}^\dagger_\mu (\bm{k})-\text{H.c.})
\right],
\end{equation}
%%%%
where the complex function 
$\alpha^\mu (\bm{k})$ characterizes the amplitude and phase of initial electromagnetic field. 
The form of the complex function 
$\alpha^\mu (\bm{k})$ is restricted by the auxiliary condition in the Becchi-Rouet-Stora-Tyutin (BRST) formalism~\cite{Sugiyama1}.
The coherent state $|\alpha\rangle_{\text{EM}}$ is interpreted as a state in which an electromagnetic field mode exists following Gauss's law due to the presence of charged particles (see Appendix A of Ref.~\cite{Sugiyama1}).
In QED, the current operator is given by the Dirac field.
Since we only focus on the localized state in the present analysis, the current operator of the field is given by the localized current of each particle, 
\begin{align}
\hat{J}^{\mu}_{j\text{I}}(x)
|R\rangle_{j}
\approx
J^{\mu}_{j{\text{R}}}(x)
|R\rangle_{j},
\quad
\hat{J}^{\mu}_{j\text{I}}(x)
|L\rangle_{j}
\approx
J^{\mu}_{j{\text{L}}}(x)
|L\rangle_{j}.
\label{approx3}
\end{align}
%where $J^{\mu}_{j{\text{R}}}(x)$ and $J^{\mu}_{j{\text{L}}}(x)$ are the current of each particle.
%The explicit forms of $J^{\mu}_{j{\text{R}}}(x)$ and $J^{\mu}_{j{\text{L}}}(x)$ are given by
The explicit forms of $J^{\mu}_{j{\text{R}}}(x)$ and $J^{\mu}_{j{\text{L}}}(x)$ are
\begin{align}
J_{j\text{R}}^{\mu}(x)
=
e_{j}\int d \tau 
\frac{d X_{j\text{R}}^{\mu}}{d \tau} \delta^{(4)}
\left(x-X_{j\text{R}}(\tau)\right)
\quad 
J_{j\text{L}}^{\mu}(x)
=e_{j} \int d \tau 
\frac{d X_{j\text{L}}^{\mu}}{d \tau} \delta^{(4)}
\left(x-X_{j\text{L}}(\tau)\right),
\end{align}
where 
$X^\mu_{j\text{R}}(\tau)$ and 
$X^\mu_{j\text{L}} (\tau)$ with $j=\text{A}, \text{B}$ representing the paths of each particle with coupling constants $e_{\text{A}}$ and $e_{\text{B}}$.
Hence, we can proceed with our computation without treating the field degrees of freedom. 
In detail, the above equations are valid~\cite{Sugiyama1, Sugiyama2, Ford1993, Ford1997, Breuer2001} when: 
\begin{quote}
 \begin{enumerate}
  \item The de Brogile wavelength is smaller than the width of the particle wavepacket.
  \item The Compton wavelength of the charged particle is much shorter than the wavelength of photon emitted from the charged particle.
  \footnote
{
Note that we have to take the coupling between the spin and external magnetic field to create a spatially superposed state of a charged particle, though we ignored the coupling between the vacuum fluctuation of the electromagnetic field, induced by the charged particle, and spin degrees of freedom. 
However, we might need to consider the coupling between the spin and the fluctuation of 
%externel
magnetic field.
The magnitude of the electric force due to vacuum fluctuation is evaluated as
\begin{align}
\Delta F
=
e\Delta E
\sim
\frac{e}{T^2},
\nonumber
\end{align}
where $T$ is the time scale of the particle in the superposition state.
On the other hand, the magnetic force due to the fluctuation, which acts a spin-$\frac{1}{2}$ particle, is estimated as
\begin{align}
\Delta F
=
\frac{1}{2}g_{{f}}\mu_{\text{B}}
\partial_{x} (\Delta B)
\sim
\frac{e}{T^2}
\nonumber
\end{align}
where $\mu_{\text{B}}=e/2m$ is the Bohr magneton with the electric charge $e$ and the mass $m$, and $g_{{f}} \sim 2$ is the electronic g-factor.
$\Delta B$ is the vacuum fluctuation of the magnetic field, which satisfies 
$\Delta B=\Delta E/c=\Delta E \sim 1/T^2$ with the natural units.
$\partial_{x} (\Delta B)$ is the field gradient in the $x$-direction, and we estimated the scale of the spatial derivative by the Compton wavelength of the charged particle $\partial_{x}\sim m$.
This means that the coupling between the vacuum fluctuation of the electromagnetic field and spin degrees of freedom could be an additional effect of the same order of the effect of the vacuum fluctuation of the magnetic force. 
}
 \end{enumerate}
\end{quote}
The first condition justifies that the state of particle is localized.
The second one neglects the process of a pair creation and annihilation. 
This means that we do not need to consider a larger Hilbert space.
The initial state evolves as follows:
%%%%
\begin{align}
|\Psi(T)\rangle 
&= 
\exp\big[-i \hat{H} T\big]|\Psi(0)\rangle
\nonumber\\
&=
e^{-i\hat{H}_0T} 
\text{T}\exp\big[-i\int_{0}^{T} dt \hat{V}_\text{I}(t) \big]
|\Psi(0)\rangle
\nonumber\\
&\approx 
e^{-i\hat{H}_0T}\frac{1}{2}
\sum_\text{P,Q=R,L}|\text{P}\rangle_{\text{A}} |\text{Q}\rangle_{\text{B}} \hat{U}_\text{PQ} |\alpha\rangle_\text{EM}
\nonumber 
\\
&=
\frac{1}{2}
\sum_\text{P,Q=R,L}|\text{P}_\text{f} \rangle_{\text{A}} |\text{Q}_\text{f} \rangle_{\text{B}} \, e^{-i\hat{H}_\text{EM} T} \hat{U}_\text{PQ} |\alpha\rangle_\text{EM},
\label{state}
\end{align}
%%%%
where $T$ is the total time scale, and particles A and B are spatially superposed.
We used the approximations provided in~\eqref{approx3} in the third line. 
$|\text{P}_\text{f}\rangle_{\text{A}}=e^{-i\hat{H}_{\text{A}}T}|\text{P}\rangle_{\text{A}}$ and $|\text{Q}_\text{f}\rangle_{\text{B}}=e^{-i\hat{H}_{\text{B}}T}|\text{Q}\rangle_{\text{B}}$ with $\text{P}, \text{Q}=\text{R}, \text{L}$ are the states of charged particles A and B, which moved along the paths P and Q, respectively.
The unitary operator $\hat{U}_{\text{PQ}}$ is expressed as
%%%%
\begin{align}
\hat{U}_{\text{PQ}} &=\text{T} \exp \left[-i \int_{0}^{T} d t \int d^{3} x\left(J_{\text{AP}}^{\mu}+J_{\text{BQ}}^{\mu}\right) \hat{A}_{\mu}^{\text{I}}(x)\right],
\end{align}
%%%%
where 
$\text{T}$ denotes the time ordering, and 
$\hat{A}^\text{I}_\mu$ is the gauge field operator in the interaction picture.

\subsection{Density matrix of charged particles and its eigenvalues}
Herein, we use the negativity $\mathscr{N}=\max[-\lambda_\text{min},0]$ to quantify the entanglement between two particles.
We derive the minimum eigenvalue $\lambda_{\text{min}}$ of the partially transposed density matrix of the particles,
$\rho^{\text{T}_\text{A}}_{\text{AB}}$.
By tracing out the degrees of freedom of the electromagnetic field to focus on the quantum state of the charged particles, we obtain the reduced density matrix of particles A and B as follows:
%%%%
\begin{align}
\rho_{\text{AB}} 
&=
\text{Tr}_{\text{EM}}[|\Psi(T)\rangle\langle\Psi(T)|]
\nonumber\\
\quad
&=
\frac{1}{4}
\sum_{\text{P}, \text{Q}=\text{R}, \text{L}}
\sum_{\text{P}', \text{Q}'=\text{R}, \text{L}}
{}_{\text{EM}}\langle\alpha|
\hat{U}^{\dagger}_{\text{P}'\text{Q}'}
\hat{U}_{\text{PQ}}
|\alpha\rangle_\text{EM}
\,
|\text{P}_\text{f}\rangle_{\text{A}}\langle {\text{P}'}_\text{f}| \otimes  
|\text{Q}_\text{f}\rangle_{\text{B}}\langle {\text{Q}'}_\text{f}|
\nonumber\\
\quad
&=
\frac{1}{4}
\sum_{\text{P}, \text{Q}=\text{R}, \text{L}}
\sum_{\text{P}', \text{Q}'=\text{R}, \text{L}}
e^{-\Gamma_{\text{P}'\text{Q}'\text{PQ}}
+i\Phi_{\text{P}'\text{Q}'\text{PQ}}}
\,
|\text{P}_\text{f}\rangle_{\text{A}}\langle {\text{P}'}_\text{f}| \otimes  
|\text{Q}_\text{f}\rangle_{\text{B}}\langle {\text{Q}'}_\text{f}|,
\end{align}
%%%%
where 
$|\text{P}_\text{f}\rangle_{\text{A}}
=e^{-i\hat{H}_\text{A}T}|\text{P}\rangle_{\text{A}}$ 
and 
$|\text{Q}_\text{f}\rangle_{\text{B}}
=e^{-i\hat{H}_\text{B}T}|\text{Q}\rangle_{\text{B}}$ 
are the states of the charged particles A and B, which move along the paths P and Q, respectively.
The quantities 
$\Gamma_{\text{P}'\text{Q}'\text{PQ}}$ and $\Phi_{\text{P}'\text{Q}'\text{PQ}}$ are expressed as
%%%%
\begin{align}
\Gamma_{\text{P}'\text{Q}'\text{PQ}}&=\frac{1}{4}\int d^4x \int d^4y (J^\mu_{\text{P}'\text{Q}'}(x)-J^\mu_{\text{PQ}}(x))
(J^\nu_{\text{P}'\text{Q}'}(y)-J^\nu_{\text{PQ}}(y))
\langle \bigl\{\hat{A}^\text{I}_\mu (x), \hat{A}^\text{I}_\nu (y)\bigr\}\rangle,
\\
\Phi_{\text{P}'\text{Q}'\text{PQ}}
&=
\int d^4x (J^\mu_{\text{P}'\text{Q}'}(x)-J^\mu_{\text{PQ}}(x))A_\mu (x)-\frac{1}{2}\int d^4x \int d^4y (J^\mu_{\text{P}'\text{Q}'}(x)-J^\mu_{\text{PQ}}(x))(J^\nu_{\text{P}'\text{Q}'}(y)+J^\nu_{\text{PQ}}(y))G^\text{r}_{\mu \nu} (x,y).
\label{PhiP'Q'PQ}
\end{align}
The field $A_\mu (x)$ in \eqref{PhiP'Q'PQ} is given by
%%%%
\begin{equation}
 A_\mu (x)= \int \frac{d^3k}{(2\pi)^{3/2} \sqrt{2k^0}}( \alpha_\mu (\bm{k})e^{ik_\nu x^\nu}+ c.c.),
\end{equation}
%%%% 
and the complex function $\alpha_{\mu}(\bm{k})$ satisfies 
%%%%
\begin{equation}
k^{\mu} \alpha_{\mu}(\bm{k})=-\frac{\tilde{J}^{0}(\bm{k})}{\sqrt{2k^{0}}}
\end{equation}
%%%%
to ensure the BRST condition (see Appendix in \cite{Sugiyama1}).  
$\tilde{J}^0(\bm{k})=\tilde{J}^0_\text{A}(\bm{k})+\tilde{J}^0_\text{B}(\bm{k})$
is the eigenvalue of the Fourier transform of the charged current $\hat{\tilde{J}}^0(\bm{k})=\hat{\tilde{J}}^0_\text{A}(\bm{k})+\hat{\tilde{J}}^0_\text{B}(\bm{k})$ at the initial time 
$t=0$.
The density matrix $\rho_{\text{AB}}$ is decomposed by using the unitary operator $\mathcal{U}$ as follows:
\begin{align}
\rho_{\text{AB}}
&=
\mathcal{U}\tilde{\rho}_{\text{AB}}\mathcal{U}^{\dagger}
=
\begin{pmatrix}
U&&\huge{O}\\
\huge{O}&&U
\end{pmatrix}
\begin{pmatrix}
W&&Z\\
Z^{\dagger}&&W
\end{pmatrix}
\begin{pmatrix}
V&&\huge{O}\\
\huge{O}&&V
\end{pmatrix},
\end{align}
where we defined $2\times2$ matrices $W$ and $Z$  as
\begin{align}
W=
\frac{1}{4}
\begin{pmatrix}
1&&e^{-\Gamma^{\text{EM}}_{\text{B}}}\\
e^{-\Gamma^{\text{EM}}_{\text{B}}}&&1
\end{pmatrix}
,
\quad
Z=
\frac{1}{4}
\begin{pmatrix}
e^{-\Gamma_{\text{A}}+i\Theta^{\text{EM}}}
&&
e^{-(\Gamma^{\text{EM}}_{\text{A}}+\Gamma^{\text{EM}}_{\text{B}}+\Gamma^{\text{EM}}_{\text{c}})}
\\
e^{-(\Gamma^{\text{EM}}_{\text{A}}+\Gamma^{\text{EM}}_{\text{B}}-\Gamma^{\text{EM}}_{\text{c}})}
&&
e^{-\Gamma_{\text{A}}-i\Theta^{\text{EM}}}
\end{pmatrix}
\end{align}
with
\begin{align}
U=
\begin{pmatrix}
e^{\frac{i}{2}(\int d^4x d^4y \Delta J^{\mu}_{\text{B}}J^{\nu}_{\text{AR}}G^{r}_{\mu\nu}-\Phi^{\text{EM}}_{\text{B}})}&&0\\
0&&e^{-\frac{i}{2}(\int d^4x d^4y \Delta J^{\mu}_{\text{B}}J^{\nu}_{\text{AR}}G^{r}_{\mu\nu}-\Phi^{\text{EM}}_{\text{B}})}
\end{pmatrix},
\end{align}
and
\begin{align}
V=
\begin{pmatrix}
e^{\frac{i}{2}(\int d^4x d^4y \Delta J^{\mu}_{\text{B}}J^{\nu}_{\text{AL}}G^{r}_{\mu\nu}-\Phi^{\text{EM}}_{\text{B}})}&&0\\
0&&e^{-\frac{i}{2}(\int d^4x d^4y \Delta J^{\mu}_{\text{B}}J^{\nu}_{\text{AL}}G^{r}_{\mu\nu}-\Phi^{\text{EM}}_{\text{B}})}
\end{pmatrix}
.
\end{align}
Here, the quantities
$\Gamma^{\text{EM}}_i$ $(i=\text{A}, \text{B})$ and $\Gamma^{\text{EM}}_{\text{c}}$ are introduced in Eqs.~\eqref{gammaiEM} and ~\eqref{gammacEM}, respectively, and $\Phi^{\text{EM}}_\text{B}$ is expressed as
\begin{align}
\Phi^{\text{EM}}_\text{B}
&=
\int d^4x \Delta J^{\mu}_{\text{B}}(x)A_{\mu}(x)
-\frac{1}{2}\int d^4x d^4y\Delta J^{\mu}_{\text{B}}(x)(J^{\nu}_{\text{BR}}(y)+J^{\nu}_{\text{BL}}(y))G^{\text{r}}_{\mu\nu}(x,y).
\end{align}
In particular, the quantity $\Theta^{\text{EM}}$ is related to $\Phi^{\text{EM}}_{\text{AB}}$ and $\Phi^{\text{EM}}_{\text{BA}}$ as follows:
\begin{align}
\Theta^{\text{EM}}
=
-\frac{i}{2}\int d^4x \int d^4y 
\Delta J^{\mu}_{\text{A}}(x)\Delta J^{\nu}_{\text{B}}(y)
[\hat{A}^\text{I}_\mu (x), \hat{A}^\text{I}_\nu (y)]
=
-\frac{i}{2}
[\hat{\phi}^{\text{EM}}_{\text{A}}, \hat{\phi}^{\text{EM}}_{\text{B}}]
\end{align}
with
\begin{align}
[\hat{\phi}^{\text{EM}}_{\text{A}}, \hat{\phi}^{\text{EM}}_{\text{B}}]
&=
\int d^4x d^4y \Delta J^{\mu}_{\text{A}}(x)\Delta J^{\nu}_{\text{B}}(y)
[\hat{A}^{\text{I}}_{\mu}(x), \hat{A}^{\text{I}}_{\nu}(y)]
\nonumber\\
\quad
&=
\int d^4x d^4y \Delta J^{\mu}_{\text{A}}(x)\Delta J^{\nu}_{\text{B}}(y)
[\hat{A}^{\text{I}}_{\mu}(x), \hat{A}^{\text{I}}_{\nu}(y)]\theta (x^{0}-y^{0})
\nonumber\\
\quad
&+
\int d^4x d^4y \Delta J^{\mu}_{\text{A}}(x)\Delta J^{\nu}_{\text{B}}(y)
[\hat{A}^{\text{I}}_{\mu}(x), \hat{A}^{\text{I}}_{\nu}(y)]\theta (y^{0}-x^{0})
\nonumber\\
\quad
&=
i\int d^4x\Delta J^{\mu}_{\text{A}}(x)\Delta J^{\nu}_{\text{B}}(y)
G^{\text{r}}_{\mu\nu}(x,y)
-i\int d^4x\Delta J^{\mu}_{\text{B}}(x)\Delta J^{\nu}_{\text{A}(y)}
G^{\text{r}}_{\mu\nu}(x,y)
\nonumber\\
\quad
&=
i(\Phi^{\text{EM}}_{\text{AB}}-\Phi^{\text{EM}}_{\text{BA}}),
\label{commutation}
\end{align}
where we inserted the step functions $\theta (x^{0}-y^{0})+\theta (y^{0}-x^{0})$ in the second line and changed the variable $x^{\mu}$ to $y^{\mu}$ and indices ${\mu}$ to ${\nu}$ of the second term in the third line.
In the fourth line, the retarded Green's function $G^{\text{r}}_{\mu\nu}(x,y)=-i[\hat{A}^\text{I}_\mu (x), \hat{A}^\text{I}_\nu (y)]\theta(x^{0}-y^{0})$ was used.
Because the entanglement is invariant under a unitary operation, $\mathcal{U}$, the eigenvalue of $\rho_{\text{AB}}^{\text{T}_{\text{A}}}$ and $\tilde{\rho}_{\text{AB}}^{\text{T}_{\text{A}}}$ are equal.
Thus, we obtain the following eigenvalues
\begin{align}
\lambda_{\pm}
&[\rho_\text{AB}^{\text{T}_{\text{A}}}]
\nonumber\\
\quad
&=
\frac{1}{4}\Big[1-e^{-\Gamma^{\text{EM}}_{\text{A}}-\Gamma^{\text{EM}}_{\text{B}}} \cosh[\Gamma^{\text{EM}}_\text{c}]\pm \Big\{\big(e^{-\Gamma^{\text{EM}}_{\text{A}}}-e^{-\Gamma^{\text{EM}}_{\text{B}}}\big)^2+4e^{-\Gamma^{\text{EM}}_{\text{A}}-\Gamma^{\text{EM}}_{\text{B}}}\sin^2\Big[\frac{\Phi^{\text{EM}}}{2}\Big]+e^{-2\Gamma^{\text{EM}}_{\text{A}}-2\Gamma^{\text{EM}}_{\text{B}}} \sinh^2[\Gamma^{\text{EM}}_\text{c}]\Big\}^{\frac{1}{2}}\Big],
\\
\quad
\lambda'_{\pm}
&[\rho_{\text{AB}}^{\text{T}_{\text{A}}}]
\nonumber\\
\quad
&=
\frac{1}{4}\Big[1+e^{-\Gamma^{\text{EM}}_{\text{A}}-\Gamma^{\text{EM}}_{\text{B}}} \cosh[\Gamma^{\text{EM}}_\text{c}]\pm \Big\{\big(e^{-\Gamma^{\text{EM}}_{\text{A}}}-e^{-\Gamma^{\text{EM}}_{\text{B}}}\big)^2+4e^{-\Gamma^{\text{EM}}_{\text{A}}-\Gamma^{\text{EM}}_{\text{B}}}\cos^2\Big[\frac{\Phi^{\text{EM}}}{2}\Big]+e^{-2\Gamma^{\text{EM}}_{\text{A}}-2\Gamma^{\text{EM}}_{\text{B}}} \sinh^2[\Gamma^{\text{EM}}_\text{c}]\Big\}^{\frac{1}{2}}\Big],
\end{align}
where $\Phi^{\text{EM}}$ is expressed as in Eq.~\eqref{PhiEM}.
Note that $\lambda_{-}[\rho_{\text{AB}}^{\text{T}_{\text{A}}}]$ is the minimum eigenvalue $\lambda^{\text{EM}}_{\text{min}}$
\begin{align}
\lambda^{\text{EM}}_{\text{min}}
=
\frac{1}{4}\Big[1-e^{-\Gamma^{\text{EM}}_{\text{A}}-\Gamma^{\text{EM}}_{\text{B}}} \cosh[\Gamma^{\text{EM}}_\text{c}]-\Big\{\big(e^{-\Gamma^{\text{EM}}_{\text{A}}}-e^{-\Gamma^{\text{EM}}_{\text{B}}}\big)^2+4e^{-\Gamma^{\text{EM}}_{\text{A}}-\Gamma^{\text{EM}}_{\text{B}}}\sin^2\Big[\frac{\Phi^{\text{EM}}}{2}\Big]+e^{-2\Gamma^{\text{EM}}_{\text{A}}-2\Gamma^{\text{EM}}_{\text{B}}} \sinh^2[\Gamma^{\text{EM}}_\text{c}]\Big\}^{\frac{1}{2}}\Big].
\label{negativityEM}
\end{align}
The gravitational version of the minimum eigenvalue $\lambda^{\text{GR}}_{\text{min}}$ can be obtained by replacing $\Gamma^{\text{EM}}_{i}$, $\Gamma^{\text{EM}}_{\text{c}}$, and $\Phi^{\text{EM}}$ in the result above~\eqref{negativityEM} with $\Gamma^{\text{GR}}_{i}$, $\Gamma^{\text{GR}}_{\text{c}}$, and $\Phi^{\text{GR}}$, respectively, as shown:
\begin{align}
\lambda^{\text{GR}}_{\text{min}}
=
\frac{1}{4}\Big[1-e^{-\Gamma^{\text{GR}}_{\text{A}}-\Gamma^{\text{GR}}_{\text{B}}} \cosh[\Gamma^{\text{GR}}_\text{c}]-\Big\{\big(e^{-\Gamma^{\text{GR}}_{\text{A}}}-e^{-\Gamma^{\text{GR}}_{\text{B}}}\big)^2+4e^{-\Gamma^{\text{GR}}_{\text{A}}-\Gamma^{\text{GR}}_{\text{B}}}\sin^2\Big[\frac{\Phi^{\text{GR}}}{2}\Big]+e^{-2\Gamma^{\text{GR}}_{\text{A}}-2\Gamma^{\text{GR}}_{\text{B}}} \sinh^2[\Gamma^{\text{GR}}_\text{c}]\Big\}^{\frac{1}{2}}\Big],
\label{negativityg}
\end{align}
%%%%
where $\Gamma^{\text{GR}}_{i}$, $\Gamma^{\text{GR}}_{\text{c}}$, and $\Phi^{\text{GR}}$ are defined as shown in Eqs.~\eqref{gammaig}, \eqref{gammacg}, and \eqref{phig}.
The results of Eqs.~\eqref{negativityEM} and \eqref{negativityg} are extended as $\lambda_{\text{min}}$ presented as Eq.~\eqref{negativity} herein.

\subsection{Complementarity inequality in QED}
Here, we present the QED results for the complementarity inequality.
First, we computed the visibility ($\mathcal{V}^{\text{EM}}_{\text{A}}$)
and the distinguishability 
($\mathcal{D}^{\text{EM}}_{\text{B}}$).
The visibility $\mathcal{V}^{\text{EM}}_{\text{A}}$ describes the extent to which the coherence of charged particle A remains when Alice performs an interference experiment.
The distinguishability $\mathcal{D}^{\text{EM}}_{\text{B}}$ characterizes how particle B can distinguish the path of particle A from the state of particle B.
The visibility $\mathcal{V}^{\text{EM}}_{\text{A}}$ of charged particle A is expressed as
\begin{align}
\mathcal{V}^{\text{EM}}_{\text{A}}
=2|_{\text{A}}\langle \text{L}_{\text{f}}|\rho^{\text{EM}}_{\text{A}}|\text{R}_{\text{f}}\rangle_{\text{A}}|.
\label{defvis}
\end{align}
The quantum state of particle A $\rho_{\text{A}}$ is obtained by tracing out the degrees of freedom of particle B and the electromagnetic field:
\begin{align}
\rho_{\text{A}}
&=
\text{Tr}_{\text{B}, \text{EM}}[|\Psi(T)\rangle \langle\Psi(T)|]
\nonumber\\
\quad
&=
\frac{1}{2}
\begin{pmatrix}
1&\frac{1}{2}e^{-\Gamma^{\text{EM}}_{\text{A}}+i\Phi^{\text{EM}}_\text{A}}\Big(e^{-i\int d^4x \Delta J^{\mu}_{\text{A}}(x) A_{\text{BR} \mu}(x)}+e^{-i\int d^4x\Delta J^{\mu}_{\text{A}}(x) A_{\text{BL} \mu}(x)}\Big)
\\
\quad * \quad &1
\end{pmatrix}
,
\label{rhoa}
\end{align}
where we used the basis 
$\{|\text{R}_\text{f} \rangle_\text{A}, |\text{L}_\text{f} \rangle_\text{A} \}$ to represent the density matrix, and 
$*$ is the complex conjugate of the 
$(\text{R},\text{L})$ component. 
$A_{i \text{P}}^{\mu}(x)=\int d^{4} y G_{\mu \nu}^{r}(x, y) J_{i \text{P}}^{\nu}(y)$ is the retarded potential with $i=\text{A}, \text{B}$ and $\text{P}=\text{R}, \text{L}$.
The quantity $\Phi^{\text{EM}}_\text{A}$ is expressed as
\begin{align}
\Phi^{\text{EM}}_\text{A}
&=
\int d^4x \Delta J^{\mu}_{\text{A}}(x) A_{\mu}(x)
-\frac{1}{2}\int d^4x d^4y\Delta J^{\mu}_{\text{A}}(x) (J^{\nu}_{\text{AR}}(y)+J^{\nu}_{\text{AL}}(y))G^{\text{r}}_{\mu\nu}(x,y).
\end{align}
From the definition of the visibility~\eqref{defvis} and~\eqref{rhoa}, we obtain the visibility as follows: 
\begin{align}
\mathcal{V}^{\text{EM}}_{\text{A}}
=e^{-\Gamma^{\text{EM}}_{\text{A}}}
\left|
\cos
\Big(
\frac{\Phi^{\text{EM}}_{\text{AB}}}{2}
\Big)
\right|.
\label{visEM}
\end{align}

Next, we compute the distinguishability $\mathcal{D}^{\text{EM}}_{\text{B}}$.
The definition of the distinguishability $\mathcal{D}^{\text{EM}}_{\text{B}}$ is expressed as
\begin{align}
\mathcal{D}^{\text{EM}}_{\text{B}}
=
\frac{1}{2}\text{Tr}_{\text{B}}|\rho_{\text{BR}}-\rho_{\text{BL}}|,
\end{align}
where we defined $\rho_{\text{BP}}=\text{Tr}_{\text{EM}}[|\Omega_\text{P}\rangle_\text{B,EM} \langle\Omega_\text{P}|]$ with $\text{P}=\text{R}, \text{L}$ and,
$\text{Tr}|\hat{O}|=\sum_i |\lambda_i|$ is given by the eigenvalues $\lambda_i$ of a Hermitian operator $\hat{O}$.
The density operator 
$\rho_\text{BP}$ characterizes the state of particle B when particle A moves along the path P.
The vector 
$|\Omega_\text{P}\rangle_\text{B,EM}$ describes 
the composite state of particle B and the electromagnetic field when particle A moves along the path P and is introduced by rewriting the state~\eqref{state} as
\begin{align}
|\Psi(T)\rangle
&=
\frac{1}{2}
\sum_\text{P,Q=R,L}|\text{P}_\text{f} \rangle_{\text{A}} |\text{Q}_\text{f} \rangle_{\text{B}} \, e^{-i\hat{H}_\text{EM} T} \hat{U}_\text{PQ} |\alpha\rangle_\text{EM}
\nonumber\\
\quad
&
=
\frac{1}{\sqrt{2}}
|\text{R}_\text{f}\rangle_{\text{A}}|\Omega_{\text{R}}\rangle_\text{B,EM}
+
\frac{1}{\sqrt{2}}
|\text{L}_\text{f}\rangle_{\text{A}}|\Omega_{\text{L}}\rangle_\text{B,EM},
\label{stateA}
\end{align}
where we defined
\begin{align}
|\Omega_{\text{P}}\rangle_\text{B,EM}
=
\frac{1}{\sqrt{2}}
\sum_{\text{Q}=\text{R},\text{L}}|\text{Q}_{\text{f}}\rangle_{\text{B}}
e^{-i\hat{H}_\text{EM} T}\hat{U}_{\text{PQ}}|\alpha\rangle_{\text{EM}}.
\end{align}
The eigenvalues of the density matrix $\rho_{\text{BR}}-\rho_{\text{BL}}$ are
\begin{align}
\lambda_{\pm}
&=\pm\frac{1}{2}\left|e^{-\Gamma^{\text{EM}}_\text{B}+i\Phi^{\text{EM}}_\text{B}-i\int d^4x\Delta J^{\mu}_{\text{B}}A_{\text{R}\mu}}-e^{-\Gamma^{\text{EM}}_\text{B}+i\Phi^{\text{EM}}_\text{B}-i\int d^4x\Delta J^{\mu}_{\text{B}}A_{\text{L}\mu}}\right|
\nonumber\\
&
=\pm e^{-\Gamma^{\text{EM}}_{\text{B}}}\left|\sin\Big(\frac{1}{2}\int d^4x\Delta J^{\mu}_{\text{B}}(x)\Delta A_{\text{A}\mu}(x)\Big)\right|.
\end{align}
Thus, the distinguishability is expressed as
\begin{align}
\mathcal{D}^{\text{EM}}_{\text{B}}
=
\frac{1}{2}(|\lambda_{+}|+|\lambda_{-}|)
=
e^{-\Gamma^{\text{EM}}_{\text{B}}}
\left|
\sin\Big(\frac{\Phi^{\text{EM}}_\text{BA}}{2}\Big)
\right|.
\label{disEM}
\end{align}
According to Refs.~\cite{Jaeger, Englert}, there is a trade-off relationship between the visibility $\mathcal{V}^{\text{EM}}_{\text{A}}$ and the distinguishability $\mathcal{D}^{\text{EM}}_{\text{B}}$, as indicated by the following inequality:
\begin{align}
(\mathcal{V}^{\text{EM}}_{\text{A}})^2+(\mathcal{D}^{\text{EM}}_{\text{B}})^2
\leq 1.
\end{align}
Therefore, the electromagnetic version of the complementarity inequality is written as
\begin{align}
(\mathcal{V}^{\text{EM}}_{\text{A}})^2+(\mathcal{D}^{\text{EM}}_{\text{B}})^2
=
e^{-2\Gamma^{\text{EM}}_{\text{A}}}\cos^2\left(\frac{\Phi^{\text{EM}}_{\text{AB}}}{2}\right)
+
e^{-2\Gamma^{\text{EM}}_{\text{B}}}\sin^2\left(\frac{\Phi^{\text{EM}}_{\text{BA}}}{2}\right)
\label{inequalityEM}
\leq 1. 
\end{align}
By replacing the quantities $\Gamma^{\text{EM}}_{i}$, $\Gamma^{\text{EM}}_{\text{c}}$, and $\Phi^{\text{EM}}$ in the above result with $\Gamma^{\text{GR}}_{i}$, $\Gamma^{\text{GR}}_{\text{c}}$, and $\Phi^{\text{GR}}$, respectively; the inequality~\eqref{inequalityEM} in the gravitational version is written as
\begin{align}
(\mathcal{V}^{\text{GR}}_{\text{A}})^2+(\mathcal{D}^{\text{GR}}_{\text{B}})^2
=
e^{-2\Gamma^{\text{GR}}_{\text{A}}}\cos^2\left(\frac{\Phi^{\text{GR}}_{\text{AB}}}{2}\right)
+
e^{-2\Gamma^{\text{GR}}_{\text{B}}}\sin^2\left(\frac{\Phi^{\text{GR}}_{\text{BA}}}{2}\right)
\leq 1. 
\label{inequalityg}
\end{align}

\section{Demonstration of the relationship expressed in the relationship~\eqref{result}}
In this appendix, the relationship expressed in relationship~\eqref{result} is demonstrated in a numerical manner. 
For convenience, we rewrite $\lambda_\text{min}$ as 
\begin{align}
\lambda_{\text{min}}
&=
\frac{1}{4}
\Big[
1-e^{-\Gamma_{\text{A}}-\Gamma_{\text{B}}} \cosh[\Gamma_\text{c}]- 
\Big\{
\big(e^{-\Gamma_{\text{A}}}-e^{-\Gamma_{\text{B}}}\big)^2+4e^{-\Gamma_{\text{A}}-\Gamma_{\text{B}}}\sin^2\Big[\frac{(\Phi_{\text{AB}}+\Phi_{\text{BA}})}{4}\Big]+e^{-2\Gamma_{\text{A}}-2\Gamma_{\text{B}}} \sinh^2[\Gamma_\text{c}]
\Big
\}^{\frac{1}{2}}
\Big]
\nonumber\\
\quad
&=
C
\Big(
\sinh[\Gamma_{\text{A}}]\sinh[\Gamma_{\text{B}}]-\sinh^2\Big[\frac{\Gamma_{\text{c}}}{2}\Big]-\sin^2\Big[\frac{\Phi_{\text{BA}}}{4}\Big]
\Big),
\end{align}
where the coefficient 
$C$ is expressed as
\begin{align}
C=
e^{-\Gamma_{\text{A}}-\Gamma_{\text{B}}}
\Big[
1-e^{-\Gamma_{\text{A}}-\Gamma_{\text{B}}}
\cosh[\Gamma_\text{c}]+ 
\Big\{
\big(e^{-\Gamma_{\text{A}}}-e^{-\Gamma_{\text{B}}}\big)^2
+4e^{-\Gamma_{\text{A}}-\Gamma_{\text{B}}}\sin^2\Big[\frac{(\Phi_{\text{AB}}+\Phi_{\text{BA}})}{4}\Big]+e^{-2\Gamma_{\text{A}}-2\Gamma_{\text{B}}} \sinh^2[\Gamma_\text{c}]
\Big\}^{1/2}
\Big]^{-1}.
\label{coefficient}
\end{align}
The coefficient $C$ is always positive because $1-e^{-\Gamma_{\text{A}}-\Gamma_{\text{B}}}\cosh[\Gamma_\text{c}]>0$ since $\Gamma_{\text{A}}+\Gamma_{\text{B}} \geq |\Gamma_{\text{c}}|$.
Therefore, the condition $\lambda_{\text{min}}\geq0$ is equivalent to
\begin{align}
\sinh[\Gamma_{\text{A}}]\sinh[\Gamma_{\text{B}}]
-\sinh^2\Big[\frac{\Gamma_{\text{c}}}{2}\Big]-\sin^2\Big[\frac{\Phi_{\text{BA}}}
{4}\Big]
\geq0.
\label{lambdabar}
\end{align}
Hereinafter, we regard the inequality~\eqref{lambdabar} as $\lambda_{\text{min}}\geq0$ and demonstrate the relationship shown in the relationship~\eqref{result}.
The relationship~\eqref{result} can be divided into two components [\eqref{c1} and \eqref{c2}] as follows:
\begin{align}
&\Gamma_{\text{A}}\Gamma_{\text{B}}\geq\frac{\Gamma^2_{\text{c}}}{4}+\frac{\Phi^2_{\text{BA}}}{16}
\Longrightarrow
\sinh[\Gamma_{\text{A}}]\sinh[\Gamma_{\text{B}}]-\sinh^2\Big[\frac{\Gamma_{\text{c}}}{2}\Big]-\sin^2\Big[\frac{\Phi_{\text{BA}}}{4}\Big]\geq0,
\label{c1}
\\
\quad
&\sinh[\Gamma_{\text{A}}]\sinh[\Gamma_{\text{B}}]-\sinh^2\Big[\frac{\Gamma_{\text{c}}}{2}\Big]-\sin^2\Big[\frac{\Phi_{\text{BA}}}{4}\Big]\geq0
\Longrightarrow
e^{-2\Gamma_{\text{A}}}+e^{-2\Gamma_{\text{B}}}\sin^2\Big[\frac{\Phi_{\text{BA}}}{2}\Big]
\leq1.
\label{c2}
\end{align}
In the following two subsections, we examine whether the relationships above~\eqref{c1} and~\eqref{c2} are satisfied.

\subsection{Demonstration of the relationship expressed in~\eqref{c1}}
First, we demonstrate the relationship expressed in ~\eqref{c1}.
Substituting the left-hand side of the inequality expressed in~\eqref{c1} into the right-hand side, we obtain the following inequality:
\begin{align}
\sinh\big[\Gamma_{\text{A}}\big]\sinh\big[\Gamma_{\text{B}}\big]
-\sinh^2\Big[\frac{\Gamma_{\text{c}}}{2}\Big]
-\sin^2\Big(\frac{\Phi_{\text{BA}}}{4}\Big)
\geq
\sinh\big[\Gamma_{\text{A}}\big]
\sinh
\Big[
\frac{\Gamma^2_{\text{c}}}{4\Gamma_{\text{A}}}+\frac{\Phi^2_{\text{BA}}}{16\Gamma_{\text{A}}}
\Big]
-\sinh^2\Big[\frac{\Gamma_{\text{c}}}{2}\Big]
-\sin^2\Big[\frac{\Phi_{\text{BA}}}{4}\Big].
\end{align}
The goal of this subsection is to demonstrate that the right-hand side of the above inequality is always positive.
Next, we define variables 
$X_{1}:=e^{-\Gamma_{\text{A}}}, 
Y_{1}:=e^{-\Gamma^2_{\text{c}}/4\Gamma_{\text{A}}}$, 
and 
$Z_{1}:=e^{-\Phi^2_{\text{BA}}/16\Gamma_{\text{A}}}$.
Note that the ranges of $X_{1}, Y_{1}, Z_{1}$ are limited to $0<X_{1}<1$, $0<Y_{1}<1$, and $0<Z_{1}<1$, respectively.
Therefore, we compute the minimum of the following function:
\begin{align}
F(X_{1}, Y_{1}, Z_{1})
:=
\frac{1}{4}\Big(\frac{1}{X_{1}}-X_{1}\Big)\Big(\frac{1}{Y_{1}Z_{1}}-Y_{1}Z_{1}\Big)-\sinh^2\Big[\sqrt{\log X_{1}\log Y_{1}}\Big]-\sin^2\Big[\sqrt{\log X_{1}\log Z_{1}}\Big].
\end{align}
%%%%
\begin{figure}[t]
  \centering
  \begin{minipage}[b]{0.3\linewidth}
  \includegraphics[width=1\linewidth]{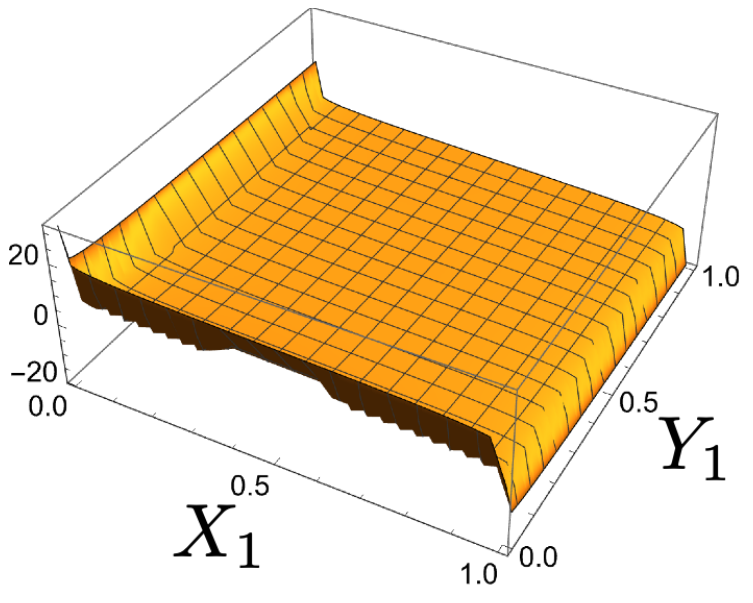}
  \end{minipage}
  \begin{minipage}[b]{0.3\linewidth}
  \includegraphics[width=1\linewidth]{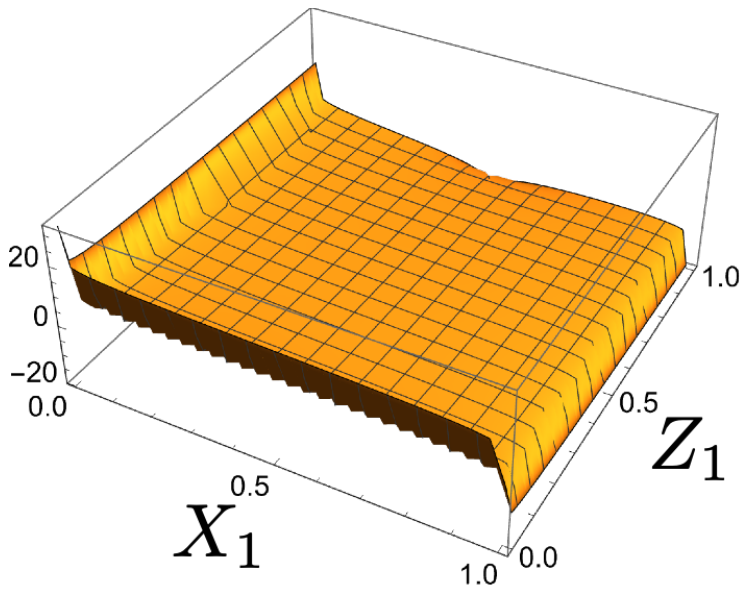}
  \end{minipage}
  \begin{minipage}[b]{0.3\linewidth}
  \includegraphics[width=1\linewidth]{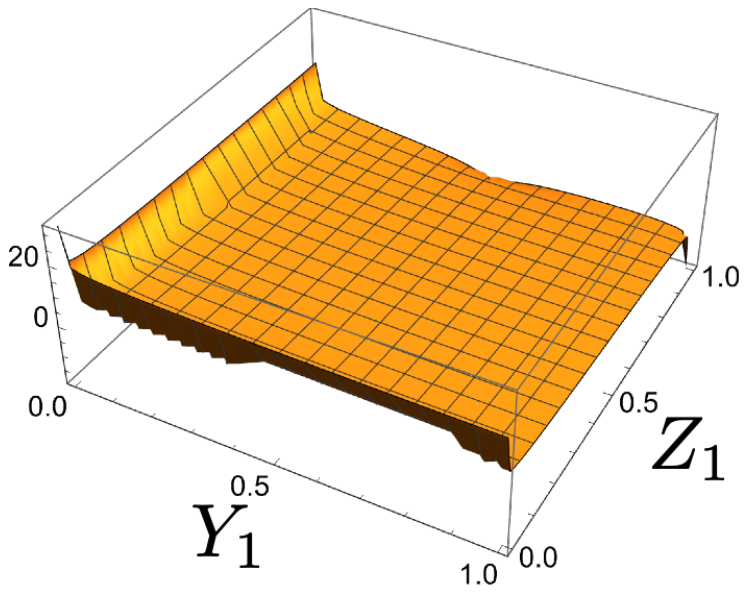}
  \end{minipage}
  \caption{Behavior of the functions of $\log [F(X_{1}, Y_{1}, 1/2)]$ (left), $\log [F(X_{1}, 1/2, Z_{1})]$ (center), and $\log [F(1/2, Y_{1}, Z_{1})]$ (right).}
  \label{fig:demo123}
\end{figure}
The function $\log[F(X_{1}, Y_{1}, Z_{1})]$ is depicted in Fig.~\ref{fig:demo123}.
The minimum value of the function $F(X_{1}, Y_{1}, Z_{1})$ is zero at 
$X_{1} =1$ based on a numerical program written using Mathematica.
These result shows that the minimum of the function $F(X_{1}, Y_{1}, Z_{1})$ is larger than zero, i.e., $F(X_{1}, Y_{1}, Z_{1})\geq0$. 
Thus, the relationship shown in ~\eqref{c1} is proven.

\subsection{Demonstration of the relationship expressed in~\eqref{c2}}
Next, we also demonstrate the relationship expressed in~\eqref{c2}.
The strategy used is the same as that used for ~\eqref{c1}, i.e., we demonstrated that the minimum of the right-most side of the inequality is greater than zero.
The left-hand-side of ~\eqref{c2} can be rewritten as
\begin{align}
\sinh[\Gamma_{\text{B}}]
\geq
\frac{\sinh^2[\Gamma_{\text{c}}/2]}{\sinh[\Gamma_{\text{A}}]}
+
\frac{\sin^2[\Phi_{\text{BA}}/4]}{\sinh[\Gamma_{\text{A}}]},
\end{align}
where $\sinh[\Gamma_{\text{A}}]>0$ because of $\Gamma_{\text{A}}>0$.
Solving the inequality above with respect to $e^{\Gamma_\text{B}}$ yields
\begin{align}
e^{\Gamma_{\text{B}}}
\geq
C+\sqrt{1+C^2},
\label{C}
\end{align}
where
$C
:=
({\sinh^2[\Gamma_{\text{c}}/2]}{\sinh[\Gamma_{\text{A}}]}
+
{\sin^2[\Phi_{\text{BA}}/4]})/{\sinh[\Gamma_{\text{A}}]}$.
Substituting the inequality in~\eqref{C} into the right-hand side of~\eqref{c2} leads to the following inequality
\begin{align}
1-e^{-2\Gamma_{\text{A}}}-e^{-2\Gamma_{\text{B}}}\sin^2\big[\frac{\Phi_{\text{BA}}}{2}\big]
\geq
1-e^{-2\Gamma_{\text{A}}}-\frac{\sin^2\big[{\Phi_{\text{BA}}}/{2}\big]}{\big(C+\sqrt{1+C^2}\big)^2}
=:
G(X_{2}, Y_{2}, Z_{2}),
\end{align}
where we defined the function $G(X_{2}, Y_{2}, Z_{2})$ as 
\begin{align}
G(X_{2}, Y_{2}, Z_{2})
:=
1-X^2_{2}-\frac{\sin^2\big[2\sin^{-1}[Z_{2}]\big]}{\big(\tilde{C}+\sqrt{1+\tilde{C}^2}\big)^2}.
\end{align}
Here, 
$X_{2}:=e^{-\Gamma_{\text{A}}}, Y_{2}:=e^{-\Gamma_{\text{c}}/2}$, $Z_{2}:=\sin[\Phi_{\text{BA}}/4]$ ($0<X_{2}<1$, $0<Y_{2}<1$, $0<Z_{2}<1$), 
and
\begin{align}
\tilde{C}
:=
\frac{\big(1/Y_{2}-Y_{2}\big)^2}{2\big(1/X_{2}-X_{2}\big)}+\frac{2Z^2_{2}}{\big(1/X_{2}-X_{2}\big)}.
\end{align}
Therefore, we focus on the minimum of the function $G(X_{2}, Y_{2}, Z_{2})$ and show that the minimum value is greater than zero.
%%%%
\begin{figure}[t]
  \centering
  \begin{minipage}[b]{0.3\linewidth}
  \includegraphics[width=1\linewidth]{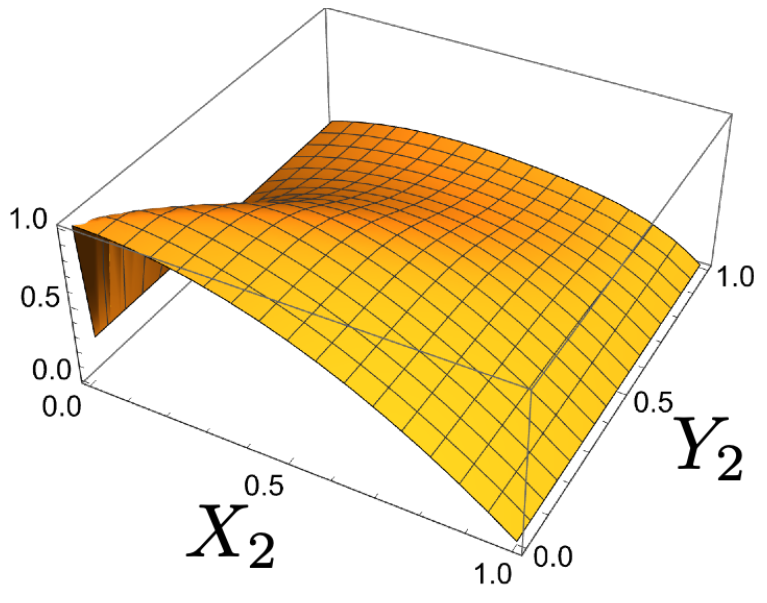}
  \end{minipage}
  \begin{minipage}[b]{0.3\linewidth}
  \includegraphics[width=1\linewidth]{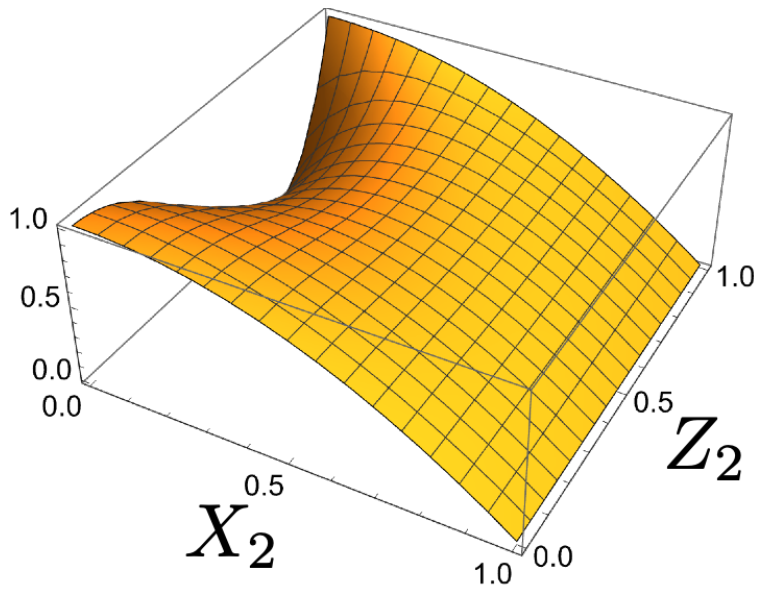}
  \end{minipage}
  \begin{minipage}[b]{0.3\linewidth}
  \includegraphics[width=1\linewidth]{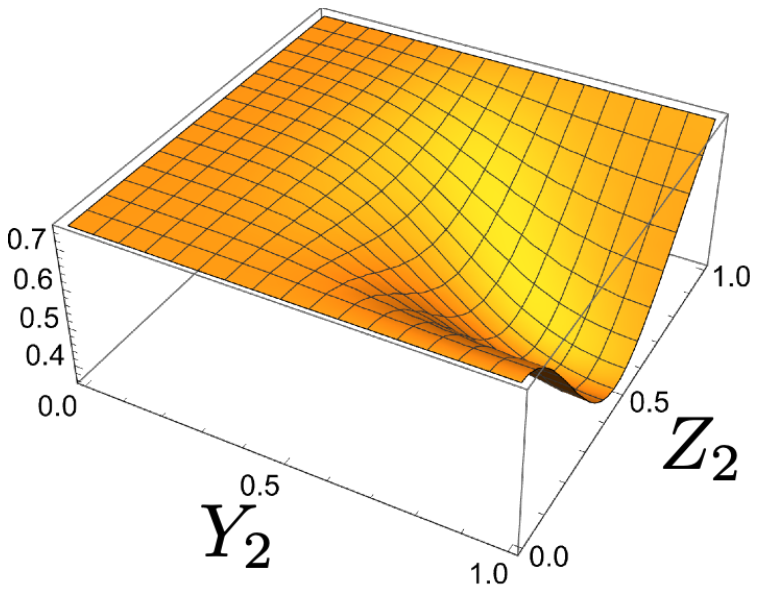}
  \end{minipage}
  \caption{Behavior of the function of $G(X_{2}, Y_{2}, 1/2)$ (left), $G(X_{2}, 1/2, Z_{2})$ (center), and $G(1/2, Y_{2}, Z_{2})$ (right).}
  \label{fig:demo456}
\end{figure}
Figure~\ref{fig:demo456} shows the behavior of the function $G(X_{2}, Y_{2}, Z_{2})$.
The minimum value of the function $G(X_{2}, Y_{2}, Z_{2})$ is zero in the limit $X_2 \rightarrow 1$ based on a  numerical program written using Mathematica; because the function $G(X_{2}, Y_{2}, Z_{2})$ is always positive, the inequality $G(X_{2}, Y_{2}, Z_{2})\geq0$ is satisfied.
Thus, the relationship expressed in~\eqref{c2} is proven.
Furthermore, based on the relationship shown in ~\eqref{c1} and~\eqref{c2}, the relationship~\eqref{result} is proven.

\end{appendix}


\begin{thebibliography}{99}
\bibitem{Bose2017} 
S. Bose, A. Mazumdar, G. W. Morley, H. Ulbricht, M Toro$\check{\text{s}}$, M. Paternostro, A. A. Geraci, P. F. Barker, M. S. Kim, and G. Milburn, 
Spin entanglement witness for quantum gravity,  Phys. Rev. Lett. {\bf 119} 240401 (2017).

\bibitem{Marlleto2017}
C. Marletto and V. Vedral, 
Gravitationally induced entanglement between two massive particles,  Phys. Rev. Lett. {\bf 119} 240402 (2017).

\bibitem{Aspelmeyer}
M. Aspelmeyer, T. J. Kippenberg, and F. Marquardt,
Cavity optomechanics,
Rev. Mod. Phys. {\bf 86}, 1391 (2014).

\bibitem{Schmole}
J. Schm\'{o}le, M. Dragosits, H. Hepach, and M. Aspelmeyer,
A micromechanical proof-of-principle experiment for measuring the gravitational force of milligram masses,
Classical Quantum Gravity {\bf 33}, 125031 (2016).

\bibitem{Blaushi} 
A. A. Balushi, W. Cong, and R. B. Mann, 
Optomechanical quantum Cavendish experiment,
Phys. Rev. A {\bf 98} 043811 (2018).

\bibitem{Christodoulou}
M. Christodoulou and C. Rovelli, 
On the possibility of laboratory evidence for quantum superposition of geometries,
Phys. Lett. B {\bf 792} 64 (2019).

\bibitem{Krisnanda}
T. Krisnanda, G. Y. Tham, M. Paternostro, and T. Paterek,
Observable quantum entanglement due to gravity,
npj Quantum Inf. {\bf 6}, 12 (2020).

\bibitem{Gnuyen}
H. Chau Nguyen and F. Bernards, 
Entanglement dynamics of two mesoscopic objects with gravitational interaction,
Eur. Phys. J. D {\bf 74}, 69 (2020).

\bibitem{Miao} 
H. Miao, D. Martynov, H. Yang, and A. Datta,
Quantum correlations of light mediated by gravity,
Phys. Rev. A {\bf 101} 063804 (2020).

\bibitem{Lopez}
S. B. Catan\~{o}-Lopez, J. G. Santiago-Condori, K. Edamatsu, and N. Matsumoto,
High-$Q$ milligram-scale monolithic pendulum for quantum-limited gravity measurements,
Phys. Rev. Lett. {\bf 124}, 221102 (2020).

\bibitem{Anastopoulos}
C. Anastopoulos and  B. L. Hu, 
Quantum superposition of two gravitational cat States,
Class. and Quant. Grav. {\bf 37} 235012 (2020).

\bibitem{Matsumura}
A. Matsumura and K. Yamamoto, 
Gravity-induced entanglement in optomechanical systems,
Phys. Rev. D {\bf 102} 106021 (2020).

\bibitem{Datta}
A. Datta and H. Miao,
Signatures of the quantum nature of gravity in the differential motion of two masses,
Quantum Sci. Technol. {\bf 6}, 045014 (2021).

\bibitem{Miki}
D. Miki, A. Matsumura, and K. Yamamoto, 
Entanglement and decoherence of massive particles due to gravity,
Phys. Rev. D {\bf 103} 026017 (2021).
			
\bibitem{Miki2}
D. Miki, A. Matsumura, and K. Yamamoto, 
Non-Gaussian entanglement in gravitating masses: The role of cumulants,
Phys. Rev. D {\bf 105}, 026011 (2022).

\bibitem{Eduardo}
E.Mart\'{i}n-Mart\'{i}nez and T.R. Perche
What gravity mediated entanglement can really tell us about quantum gravity,
arXiv:2208.09489.

\bibitem{Kaku1}
Y. Kaku, S. Maeda, Y. Nambu, and Y. Osawa,
Quantumness of gravity in harmonically trapped particles,
Phys. Rev. D {\bf 106} 126005 (2022).

\bibitem{Miki3}
D. Miki, N. Matsumoto, A. Matsumura, T. Shichijo, Y. Sugiyama, K. Yamamoto, and N. Yamamoto,
Generating quantum entanglement between macroscopic objects with continuous measurement and feedback control,
Phys. Rev. A {\bf 107}, 032410 (2023).

\bibitem{Sugiyama3}
Y. Sugiyama, T. Shichijo, N. Matsumoto, A. Matsumura, D. Miki, and K. Yamamoto,
Effective description of a suspended mirror coupled to cavity light: Limitations of Q enhancement due to normal-mode splitting by an optical spring,
Phys. Rev. A {\bf 107}, 033515 (2023).

\bibitem{Shichijo}
T. Shichijo, N. Matsumoto, A. Matsumura, D. Miki, Y. Sugiyama, and K. Yamamoto,
Quantum state of a suspended mirror coupled to cavity light
-Wiener filter analysis of the pendulum and rotational modes-,
arXiv:2303.04511.

\bibitem{Kaku2}
Y. Kaku, T. Fujita, and A. Matsumura,
Enhancement of quantum gravity signal in an optomechanical experiment,
arXiv:2306.02974.

\bibitem{Baym}
G. Baym and T. Ozawa,
Two-slit diffraction with highly charged particles: Niels bohr’s consistency argument that the electromagnetic field must be quantized, Proc. Natl. Acad. Sci. U.S.A. {\bf 106} 3035 (2009).

\bibitem{Mari}
A. Mari, G. De Palma and V. Giovannetti,
Experiments testing macroscopic quantum superpositions must be slow, Sci. Rep. {\bf 6} 22777 (2016).

\bibitem{Belenchia2018}
A. Belenchia, R. M. Wald, F. Giacomini, E. Castro-Ruiz, C. Brukner, and M. Aspelmeyer,
Quantum superposition of massive objects and the quantization of gravity, 
Phys. Rev. D {\bf 98}, 126009 (2018).

\bibitem{Belenchia2019}
A. Belenchia, R.M. Wald, F. Giacomini, E. Castro-Ruiz, v. Brukner and M. Aspelmeyer,
Information content of the gravitational field of a quantum superposition, 
Int. J. Mod. Phys. D {\bf 28} 1943001 (2019).

\bibitem{Danielson2021}
D. L. Danielson, G. Satishchandran, and R. M. Wald,
Gravitationally mediated entanglement: Newtonian field vs. gravitons, 
Phys. Rev. D {\bf 105}, 086001 (2022).

\bibitem{Pesci}
A. Pesci, 
Conditions for graviton emission in the recombination of a delocalized mass,
Quantum Rep. {\bf 5}, 426 (2023).

\bibitem{Sugiyama1}
Y. Sugiyama, A. Matsumura, and K. Yamamoto,
Effects of photon field on entanglement generation in charged particles, 
Phys. Rev. D {\bf 106}, 045009 (2022).

\bibitem{Sugiyama2}
Y. Sugiyama, A. Matsumura, and K. Yamamoto,
Consistency between causality and complementarity guaranteed by Robertson inequality in quantum field theory, 
Phys. Rev. D {\bf 106}, 125002 (2022).

\bibitem{Iso1}
Y. Hidaka, S. Iso, and K. Shimada,
Complementarity and causal propagation of decoherence by measurement in relativistic quantum field theories, 
Phys. Rev. D {\bf 106}, 076018 (2022).

\bibitem{Iso2}
Y. Hidaka, S. Iso, and K. Shimada,
Entanglement generation and decoherence
in a two-qubit system mediated by relativistic quantum field, 
Phys. Rev. D {\bf 107}, 085003 (2023). 

\bibitem{Wheeler}
Kip S. Thorne, Charles W. Misner, and John Archibald Wheeler, 
\textit{Gravitation} (Freeman, San Francisco, CA, 2018).

\bibitem{Suzuki}
F. Suzuki and F. Queisser, 
Environmental gravitational decoherence and a tensor noise model,
J. Phys. Conf. Ser. {\bf 626}, 012039 (2015).

\bibitem{Howl}
M. Christodoulou, A. Di Biagio, M. Aspelmeyer, C. Brukner, C. Rovelli, and R. Howl,
Locally mediated entanglement in linearised quantum gravity,
Phys. Rev. Lett. {\bf 130} 100202 (2023). 

\bibitem{Howl2}
See Supplemental Material at http://link.aps.org/ supplemental/10.1103/PhysRevLett.{\bf 130}.100202

\bibitem{Donoghue}
J. F. Donoghue, M. M. Ivanov, and A. Shkerin,
EPFL lectures on general relativity as a quantum field theory,
arXiv:1702.00319.

\bibitem{Jaeger}
G. Jaeger, A. Shimony, and L. Vaidman
Two interferometric complementarities, Phys. Rev. A {\bf51}, 54 (1995).

\bibitem{Englert}
B.-G. Englert,
Fringe visibility and which-Way information: An Inequality, 
Phys. Rev. Lett. {\bf 77}, 2154 (1996).

\bibitem{Vidal2002}
G. Vidal and R. F. Werner, 
Computable measure of entanglement, 
Phys. Rev. A {\bf 65}, 032314 (2002).

\bibitem{Sanpera1998}
A. Sanpera, R. Tarrach, and G. Vidal, 
Local description of quantum inseparability, 
Phys. Rev. A {\bf 58}, 826 (1998).

\bibitem{Ford1993}
L. H. Ford, 
Electromagnetic vacuum fluctuations and electron coherence, 
Phys. Rev. D {\bf 47},  5571 (1993).

\bibitem{Ford1997}
L. H. Ford, 
Electromagnetic vacuum fluctuations and electron coherence. II. Effects of wave-packet size, 
Phys. Rev. A {\bf 56}, 1812 (1997). 

\bibitem{Breuer2001}
H.-P. Breuer and F. Petruccione, 
Destruction of quantum coherence through emission of bremsstrahlung, 
Phys. Rev. A {\bf 63}, 032102 (2001).

\end{thebibliography}
\end{document}